\documentclass[preprint2]{aastex63}
\usepackage[whole]{bxcjkjatype} 
\usepackage[utf8]{inputenc}

\accepted{February 6, 2024}

\shorttitle{Formation of limb-brightened jet}
\shortauthors{Hirotani et al.}
\graphicspath{{./}{figures/}}

\begin{document}

\title{Formation of limb-brightened radio jets by angle-dependent
energy extraction from rapidly rotating black holes}

\correspondingauthor{Kouichi Hirotani, Hsien Shang}
\email{hirotani,shang@asiaa.sinica.edu}

\author[0000-0002-2472-9002]{Kouichi Hirotani}
\affiliation{Institute of Astronomy and Astrophysics, Academia Sinica, 
Taipei 10617, Taiwan, R.O.C.}

\author[0000-0001-8385-9838]{Hsien Shang （尚賢）}
\affiliation{Institute of Astronomy and Astrophysics, Academia Sinica, 
Taipei 10617, Taiwan, R.O.C.}

\author[0000-0001-5557-5387]{Ruben Krasnopolsky}
\affiliation{Institute of Astronomy and Astrophysics, Academia Sinica, 
Taipei 10617, Taiwan, R.O.C.}

\author[0000-0001-6031-7040]{Kenichi Nishikawa}
\affiliation{Department of Physics, Chemistry and Mathematics, 
Alabama A\&M University, Huntsville, AL 35811, USA
}

\begin{abstract}
By general relativistic magnetohydrodynamic simulations,
it is suggested that the rotational energy of 
a rapidly rotating black hole (BH)
is preferentially extracted
along the magnetic field lines threading the event horizon 
in the middle and lower latitudes.
Applying this angle-dependent Poynting flux to the jet downstream,
we demonstrate that the jets exhibit 
limb-brightened structures at various viewing angles,
as observed from Mrk~501, M87, and Cyg~A
between 5 and 75 degrees,
and that the limb-brightening is enhanced
when the jet is collimated strongly.
It is also found that the jet width perpendicular to 
the propagation direction shrinks at the projected distance of the altitude where the jet collimates 
from a conical shape (near the BH)
to a parabolic one (in the jet).
Comparing with the VLBI observations,
we show this collimation takes place 
within the de-projected altitude of 
$100$ Schwarzschild radii from the BH
in the case of the M87 jet.
\end{abstract}


\section{Introduction}
\label{sec:intro}
In many active galactic nuclei,
collimated jets are launched from 
their central regions,
where supermassive black holes (BHs) presumably exist.
It is widely believed that these relativistic flows
are energized either by the extraction of the BH's rotational energy
via the magnetic fields threading the event horizon
\citep{bla77} 
or by the extraction of the accretion flow's rotational energy
via the magnetic fields threading the accretion disk
\citep{blandford:1982MNRAS}.
In the recent two decades, 
general relativistic (GR) magnetohydrodynamic (MHD) simulations
successfully demonstrated that a Poynting-dominated jet can be driven by
the former process, the so-called \lq\lq Blandford-Znajek (BZ) process''
\citep{Koide:2002:Sci,mckinney:2004ApJ,komissarov:2005MNRAS,
Tchekhovskoy:2011:MNRAS,Qian:2018ApJ}.
The BZ process takes place 
when the horizon-penetrating magnetic field 
($\mbox{\boldmath$B$}$) is dragged
in the rotational direction of the BH by space-time dragging,
and when there is a meridional current 
($\mbox{\boldmath$J$}$) flowing just above the horizon
(as a part of the global pattern of magnetospheric currents).
The resultant 
$\mbox{\boldmath$J$} \times \mbox{\boldmath$B$}$ Lorentz force
exerts a counter torque on the horizon,
extracting the BH's rotational energy electromagnetically,
and carrying the energy and momentum to large distances
in the form of torsional Alfven waves
when the toroidally twisted magnetic field lines unwind.

Subsequently, applying the GR particle-in-cell (PIC) technique
to magnetically dominated BH magnetospheres,
it is confirmed that the BZ process does work
even when the Ohm's law, and hence the MHD approximation breaks down
\citep{Parfrey:2019:PhRvL,Chen:2020:ApJ,Kisaka:2020:ApJ,Crinquand:2021:AA,
Bransgrove:2021:PhRvL,Hirotani:2023:ApJ}.
However, with current computational capability,
the GRPIC technique cannot be applied to supermassive BHs,
if we consider the magnetic fields whose strengths are 
comparable to what inferred from observations,
because the plasma skin depth and/or the gyration motion
cannot be easily resolved in such cases.
As a result, GRPIC codes have been applied to 
2D (axisymmetric) magnetospheres around either
a supermassive BH with an extremely weak magnetic-field strength, or
a stellar-mass BH with a realistic magnetic-field strength.
Nevertheless, it is confirmed that the BH's rotational energy
can be efficiently extracted via the BZ procss
by both GRMHD and GRPIC techniques.

It is possible that such extracted electromagnetic energy
is converted into the kinetic and internal energies 
of electron-positron pairs at some distance from the BH,
and eventually dissipated as radiation 
via the synchrotron and inverse-Compton processes in the jet downstream.
In the present paper,
we thus semi-analytically examine the jet emission properties,
connecting the jet-launching and jet-downstream regions.

In \S~\ref{sec:BZflux},
we apply published GRMHD simulation results to the jet-launching region
near the BH,
and constrain the angular dependence of the BZ flux.
We then describe how such electromagnetic energies are converted
into the plasma's energies in the jet
in \S~\ref{sec:model},
utilizing another published MHD simulation results on jet propagation.
Then in \S~\ref{sec:jet},
we depict the expected VLBI maps from low-accretion BH systems
and demonstrate that the limb-brightened structures,
which has been observed in the jets of
Mar~501 \citep{Giroletti:2004:ApJ},
M87 \citep{Kovalev:2007:ApJL},
3C84 \citep{Nagai:2014:ApJ}, and
Cyg~A \citep{Boccardi:2016:A&A},
can be formed by virtue of the angle-dependent BZ flux.
Finally in \S~\ref{sec:disc}, we compare the present analysis
with an analytical work done by \citet{Takahashi:2018:ApJ}.

\section{Poynting flux near the Black Hole}
\label{sec:BZflux}
Let us start with constraining the meridional dependence of the BZ flux
in the BH vicinity.
In a stationary and axisymmetric BH magnetosphere,
the radial component of the BZ flux (or equivalently the Poynting flux)
is given by
\begin{equation}
  S^r= - \Omega_{\rm F} B^r B_\varphi,
\end{equation}
where 
$\Omega_{\rm F} \equiv F_{t \theta} / F_{\theta \phi}$
designates the angular frequency of the rotating magnetic field,
$B^\mu \equiv {}^\ast\! F^\mu{}_t$ does the $\mu$ component of the magnetic field;
$F$ and ${}^\ast\! F$ denote the Faraday and Maxwell tensors, respectively.
The strength of the toroidal component of the magnetic field is expressed
up to a numerical factor by the enclosed poloidal current $I(A_\varphi)$ by
\begin{equation}
  B_\varphi \approx -I(A_\varphi) / 2 \pi,
  \label{eq:Bphi}
\end{equation}
where $A_\varphi$ denotes the magnetic flux function.
In a stationary and axisymmetric magnetosphere,
a magnetic field line resides in a constant $A_\varphi$ surface.

In this paper, we model the jet flow lines in the poloidal plane
with a parabolic-like geometry 
\citep{2009ApJ...697.1164B,Takahashi:2018:ApJ},
\begin{equation}
  A_\varphi
    = A_{\rm max} 
    \left( \frac{r}{r_{\rm H}} \right)^q
    \left( 1 - \vert \cos\theta \vert \right),
  \label{eq:A3}
\end{equation}
where $r_{\rm H}= M + \sqrt{ M^2 - a^2 }$ denotes the horizon radius
in geometrized unit (i.e., $c=G=1$),
and $\theta$ does the colatitude;
$c$ and $G$ refer to the speed of light and the gravitational constant,
respectively.
Equation~(\ref{eq:A3}) gives 
a pure parabolic flowline if $q=1$
and a conical one if $q=0$.
We adopt this flow-line model,
because the M87 jet is known to be parabolic-like up to the Bondi radius
\citep{Asada:2012:ApJL}.

Differentiating equation~(\ref{eq:A3}) with respect to $\theta$, 
we obtain the radial component of the magnetic field,
\begin{equation}
  B^r= \frac{ (a-2Mr \Omega_{\rm F}) a \sin^2\theta -\Delta }
            { \Sigma^2 }
       A_{\rm max}
       \left( \frac{r}{r_{\rm H}} \right)^q,
  \label{eq:Br}
\end{equation}
where $a$ denotes the BH's spin parameter,
$\Delta \equiv r^2 -2Mr +a^2$ vanishes as the horizon, 
and $\Sigma \equiv r^2 + a^2 \cos^2\theta$.
At the horizon, $r=r_{\rm H}$,
We thus obtain the BZ flux,
\begin{equation}
  S^r_{\rm H}
    = \frac{a^2}{2\pi \Sigma^2}
      \Omega_{\rm F} ( \omega_{\rm H} - \Omega_{\rm F} ) A_{\rm max}^2
      \frac{I(A_\varphi)}{A_{\rm max} \omega_{\rm H}}
      \sin^2\theta,
  \label{eq:Sr}
\end{equation}
where $\omega_{\rm H} \equiv a/(2Mr_{\rm H})$ denotes
the angular frequency of the spinning BH.

It is reasonable to assume that the magnetic field lines are predominantly 
radial in the BH vicinity even in a magnetically dominated magnetosphere 
(in which plasma's rest-mass energy density is negligible compared
to that of the magnetic field),
because of the causality at the horizon
and the particle's inertia
\citep[e.g.,][]
{Hirotani:1992:ApJ,McKinney:2012:MNRAS,Bransgrove:2021:PhRvL}.
Then
\citet{Tchekhovskoy:2010:ApJ} performed a GRMHD simulation and demonstrated
that $\Omega_{\rm F} \approx 0.5 \omega_{\rm H}$ and
$I(A_\varphi)/(A_{\rm max}\omega_{\rm H}) 
 \approx 3 \sin((\pi/2)(A_\varphi/A_{\rm max})$
hold for a radial field line.
We thus obtain
\begin{equation}
  \frac{S^r_{\rm H}}{S^r_{0,{\rm H}}} 
  \approx 
    w(2-w) f(\theta) \cdot 3 \sin\left[ \frac{\pi}{2}(1-\cos\theta) \right]
    \sin^2\theta,
  \label{eq:Sr_2}
\end{equation}
where 
$S^r_{0,{\rm H}} \equiv 
 a^2 \omega_{\rm H}^2 A_{\rm max}^2 / ( 8 \pi \Sigma^2 )$ 
denotes the typical, angle-averaged BZ flux at the horizon,
and $w \equiv \Omega_{\rm F} / ( 0.5 \omega_{\rm H})$.
An additional factor $f=f(\theta)$ is of an order of unity,
and is set to be $f=1$ in most cases.
However, to test the effect of the vanishment of $B_\varphi$ on the equator,
we will consider the case of a vanishing 
$f(\theta)$ at $\theta= \pi/2$ in \S~\ref{sec:jet_limb}.
Nevertheless, the main conclusion of this paper,
that is, the formation of a limb-brightened jet,
is little affected by the deviation of $f$ from unity.

\begin{figure*}
\includegraphics[width=\textwidth, angle=0, scale=1.0]
{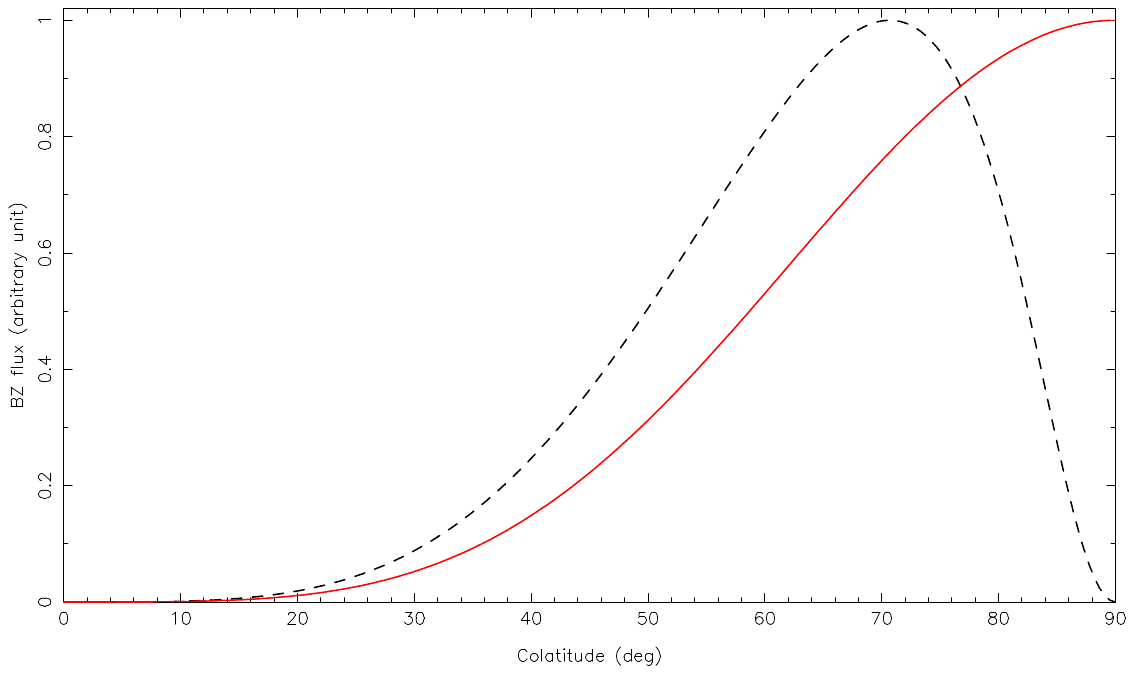}
\vspace*{-5.0truecm}
\caption{
Stationary and axisymmetric BZ flux as a function of the colatitude.
The red solid curve denotes the BZ flux when $f=1$ (see text), whereas
the black dashed one does the BZ flux 
when $f$, and hence $B_\varphi$ vanishes on the equator.
}
\label{fig:BZ_theta}
\end{figure*}

We plot $S^r_{\rm H} / S^r_{0,{\rm H}}$ 
as a function of $\theta$ in figure~\ref{fig:BZ_theta}.
The red solid curve shows the BZ flux when $f=1$.
The black dashed one does the BZ flux when
$f= x^2/(x^2+1)$, where $x= 5(\theta-\pi/2)$.
In both cases, the BZ flux is proportional to $\theta^4$ 
at $\vert \theta \vert \ll 1$.
It is, therefore, difficult to contrive a Poynting-dominated jet 
along $\theta=0$ axis within the force-free or MHD framework.
We will use the red solid case as the main scenario 
in \S~\ref{sec:jet},
and use the black dashed case only for comparison 
in \S~\ref{sec:jet_limb}

\section{Jet model}
\label{sec:model}
In this paper, 
we assume that the jet is composed of electron-positron pair plasmas,
and consider both thermal and non-thermal components.

\subsection{Kinetic flux inferred by the BZ flux}
\label{sec:jet_kinetic}
When a jet is composed of a pure pair plasma,
its kinetic flux can be computed by
\begin{equation}
  F_{\rm kin}
  = \beta c \Gamma (\Gamma-1)
    \left[ \frac32 \Theta n_\ast^{\rm th} 
           + \langle \gamma \rangle n_\ast^{\rm nt}
    \right] \cdot 2 m_{\rm e} c^2
  \label{eq:F_kin}          
\end{equation}
where 
$n_\ast^{\rm th}$ and $n_\ast^{\rm nt}$ denote
the co-moving number density of thermal and non-thermal
electrons, respectively;
$\beta c$ refers to the fluid velocity,
$\Gamma \equiv 1 / \sqrt{1-\beta^2}$ the bulk Lorentz factor,
$\Theta \equiv kT_{\rm e} / m_{\rm e} c^2$ 
the dimensionless temperature of thermal electrons,
$\langle \gamma \rangle$ 
the mean Lorentz factor of randomly moving non-thermal electrons,
and $m_{\rm e}c^2$ the electron's rest-mass energy.

Neglecting the energy dissipation in the jet
\citep{Celotti:1993:MNRAS:264},
we find that the summation of electromagnetic and kinetic energies
is conserved along each flux tube.
We thus obtain the kinetic-energy flux
\begin{equation}
  F_{\rm kin}(r,\theta)
  = \frac{1}{1+\sigma}
    \frac{B_{\rm p}(r,\theta)}{B_{\rm p}^{\rm (0)}} 
    F_{\rm BZ}^{\rm (0)}
  \label{eq:F_kin4}
\end{equation}
at position ($r$,$\theta$) in the jet,
where $\sigma$ denotes the magnetization parameter,
$B_{\rm p}(r,\theta)$ the poloidal magnetic field strength
at distance $r$ from the BH and at colatitude $\theta$ in the jet.
$F_{\rm BZ}^{(0)}$ denotes the Poynting flux at
the jet launching point, whereas
$B_{\rm p}^{\rm (0)}$ does the $B_{\rm p}$ at the same point.
At the jet base, we obtain
$F_{\rm BZ}^{\rm (0)}/B_{\rm p}^{\rm (0)}
 = S^r_{\rm H}/B_{\rm p}(r_{\rm H},\theta)$.
In addition,
$B_{\rm p}(r,\theta)$ can be computed
by equation~(\ref{eq:Br}),
and $S^r_{\rm H}/S^r_{0,{\rm H}}$ 
is known (\S~\ref{sec:BZflux}).
We constrain the normalization factor
$S^r_{0,{\rm H}} \propto A_{\rm max}^2$
by adjusting the integrated flux density
(at 86~GHz in the preset paper)
with the observed value.
Therefore, the right-hand side of equation~(\ref{eq:F_kin4})
is obtained, 
once $\sigma= \sigma(r,\theta)$ is specified.

In equations~(\ref{eq:F_kin}) and (\ref{eq:F_kin4}),
we will constrain $\Gamma= \Gamma(r)$ and $\sigma= \sigma(r)$
in \S\S~\ref{sec:jet_Lf}--\ref{sec:jet_sigma},
using a published MHD simulation result.
We set $\Theta$ to be $1$ or $3$ in the present paper.
As long as $\gamma_{\rm min}= O(1)$,
$\langle \gamma \rangle$ becomes of order of unity.
For a general argument of $\langle \gamma \rangle$ computation,
see e.g., \S~3.2 of \citet{Hirotani:2005:ApJ}.

\subsection{Emission and absorption coefficients}
\label{sec:jet_emissivity}
Let us consider how to constrain the co-moving densities,
$n_\ast^{\rm th}$ and $n_\ast^{\rm nt}$, 
combining equations~(\ref{eq:F_kin}) and (\ref{eq:F_kin4}).
For thermal pairs,
the emission coefficient is given by
\citet{Leung:2011:ApJ},
and the absorption coefficient can be computed by the Kirchhoff's law.
Their distribution function obeys
the Maxwell-J$\ddot{\rm u}$ttner distribution,
which is characterized by $\Theta$.
For non-thermal pairs, 
emission and absorption coefficients are given 
in e.g., 
\citet{Rybicki:1986:rpa, LeRoux:1961, Ginzburg:1965}.
We present the explicit expressions of these coefficients
in the Appendix.

To compute the emission and absorption coefficients,
it is essential to constrain both
$n_\ast^{\rm th}$ and $n_\ast^{\rm nt}$.
Let us parameterize the former by
$n_\ast^{\rm th}= \zeta n_\ast^{\rm nt}$,
which gives the total proper density 
$n_\ast=(1+\zeta) n_\ast^{\rm nt}$.
\citet{Ito:2008:ApJ:685} investigated 
the kinetic powers and ages of four Fanaroff-Riley II radio galaxies,
and derived $4 < 1 + \zeta < 310$
under a minimum-energy condition 
\citep[e.g.,][]{Miley:1980:ARAA, Kellermann:1981:ARAA}.

\subsection{Assumptions}
\label{sec:jet_assumptions}
For thermal pairs,
we assume that they
have a semi-relativistic temperature, $\Theta= 1.0$,
and dominate the non-thermal ones by a factor 
$\zeta=(1-0.007)/0.007 \approx 142$;
that is, $0.7$~\% of the pairs are non-thermal.
It is noteworthy that $\zeta$ merely changes 
the normalization of the jet luminosity.
Thus, when $\zeta$ has a greater value, for instance,
we can obtain a very similar jet solution 
by adopting an appropriately smaller value of $A_{\rm max}$.
In another word, the main conclusions of the present paper
--- limb brightening and constriction of jets ---
are little affected by the choice of such parameters. 

For the non-thermal pairs, 
we set the power-law index 
$p=3.0$, which gives the spectral index of 
$\alpha=-1.0$
in the optically thin frequency regime.
The lower and upper cut-off Lorentz factors are
$\gamma_{\rm min}=1$ and 
$\gamma_{\rm max}=10^5$.

We assume an energy equipartition between
the energy density of (thermal+non-thermal) pairs
and that of the random magnetic field, $B_\ast^2/8\pi$.

For the jet geometry,
we model the jet flow lines so that they
may coincide with the magnetic-flux surface of 
a parabolic-like geometry (eq.~[\ref{eq:A3}]).
In a super-fast-magnetosonic region,
magnetic field lines are substantially wound up in the trailing direction
\citep{came86b};
accordingly, the motion of an MHD flow becomes predominantly poloidal
far outside the outer light surface
\citep{takahashi:1990ApJ,toma2013PTEP}.
Thus, we neglect the rotational motion of fluids 
around the jet axis.

In this paper, we assume that 
$q$ (eq.~[\ref{eq:A3}]) increases linearly with $r$
inside the collimation altitude $r=r_{\rm c}$,
and parameterise it by
\begin{equation}
  q
  = q_\infty \cdot
    {\rm min} \left( \frac{r-r_{\rm H}}{r_{\rm c}} , 1 \right) .
  \label{eq:nu_evolv}
\end{equation}
Accordingly, the flow lines change
from a conical shape (i.e., $q=0$) 
near the BH at $r-r_{\rm H} \ll r_{\rm c}$
to a parabolic one (i.e., $q= q_\infty$) 
beyond the altitude $r = r_{\rm c}$.
We adopt an observational value, $q_\infty= 0.75$
\citep{Nakamura:2018:ApJ}.

As an example, we apply the present method to the M87 jet,
and adopt a BH mass $6.4 \times 10^9 M_\odot$ 
\citep{Gebhardt:2011:ApJ,EventHorizonTelescopeCollaboration:2019:ApJL,
Liepold:2023:ApJL}
and the luminosity distance $16.7$~Mpc.
An angular scale of $1$~mas corresponds to a spatial scale 
$140 R_{\rm S}$,
where $R_{\rm S} \equiv 2GM c^{-2}= 2M$
denotes the Schwarzschild radius. 
We assume a rapidly rotating BH
and adopt $a=0.9M$
\citep{Li:2009:ApJ,Dokuchaev:2023:Astro,Feng:2017:MNRAS}.

\subsection{Constraining the jet kinematics}
\label{sec:jet_kinematics}
Analyzing multi-epoch VLBI data 
with the wavelet-based image segmentation and evaluation method,
\citet{Mertens:2016:AA} investigated the kinematics of the M87 jet.
They revealed a stratification of the jet flow with a slow, subluminal 
and fast, superluminal (i.e., relativistic) velocity components.
Since the fast velocity component is present in all the three
emission ridges, they argue that it can be associated with the
bulk fluid motion in the sheath, rather than the spine of the jet.
On the other hand, the slow component may be interpreted as a pattern speed 
associated with an instability.

On these grounds, we consider that the fast, relativistic velocity component
represents the plasma motion in the jet.

\subsubsection{Evolution of the bulk Lorentz factor}
\label{sec:jet_Lf}
Using the observed apparent velocity as a function of distance,
\citet{Mertens:2016:AA} revealed that 
the bulk Lorentz factor increases 
from $\Gamma \approx 2$ at distance $\varpi \approx 0.4$~mas from the jet axis
(i.e., at de-projected distance $r \approx 1$~mas from the BH) to 
$\Gamma \approx 4.5$ at $\varpi \approx 1.5$~mas
(i.e., at $r \approx 20$~mas).
It should be noted that $r$ in \citet{Mertens:2016:AA} designates
the distance $\varpi$ from the jet axis in the present notation.
We thus approximate the Lorentz factor evolution by
\begin{equation}
  \Gamma
  = 2 + 2 \min\left[ \log_{10} \left( \frac{r}{\rm mas} \right) , 2 \right]
  \label{eq:Lf_evol}
\end{equation}
in the present paper.
At larger scales, $r \gg 1$~mas,
the bulk Lorentz factor is found to be $\Gamma \sim 6$
by HST observations \citep{Biretta:1999:ApJ}.
We thus set an upper limit of $6$ in the right-hand side.
Nevertheless, the saturation of $\Gamma$ at $6$ does not
play a role in the present analysis,
because we consider only the inner jet,
$r \ll 100$~mas, in the de-projected distance from the BH.

\subsubsection{Evolution of the magnetization parameter}
\label{sec:jet_sigma}
In relativistic MHD jet models,
most of the conversion of energy from Poynting to kinetic takes place
when highly compressed magnetic field lines (in the upstream)
expands in the super-magnetosonic region
\citep{Li:1992:ApJ,Chiueh:1998:ApJ,Vlahakis:2003:ApJ}.
Accordingly, the increased plasma's mass winds the field lines
into the counter-rotational direction to collimate the field lines
into a parabolic geometry due to the hoop stress
\citep{Tomimatsu:2003:ApJ}.
The fast-magnetosonic point is located typically
at several light cylinder radii, $\varpi_{\rm LC}$
\citep{Li:1992:ApJ,Chiueh:1998:ApJ},
whereas $\varpi_{\rm LC}$
typically becomes a few Schwarzschild radii
\citep{Znajek:1977:MNRAS,takahashi:1990ApJ}.
For instance,
assuming a quasi-parabolic jet flow line, and
applying a cold, ideal MHD model to the M87 jet,
\citet{Mertens:2016:AA} finds that
the magnetization parameter $\sigma$,
the ratio between the Poynting and kinetic energy fluxes,
is much greater than $1.0$ near the BH,
but decreases with distance from the BH
to become $2.0$ at fast point, 
whose projected distance is 
$z \equiv r \sin\theta_{\rm v} \approx 1$~mas 
along the jet from the BH,
where $\theta_{\rm v}$ denotes the observer's viewing angle
of the jet.
In what follows, 
$z$ designates the ordinate of the expected VLBI maps
(figs.~\ref{fig:map_collimation}, \ref{fig:map_angle}
 \ref{fig:map_Bp}, and \ref{fig:map_Bp}), 
whereas $\varpi$ does the abscissa
of figures~\ref{fig:map_collimation}--\ref{fig:slice_rc_obs}.

On these grounds,
we assume that $\sigma$ evolves with the de-projected distance $r$ 
from the BH such that
\begin{equation}
  \sigma(r)= 2 \left( \frac{r}{150 R_{\rm S}} \right)^{-1/4},
  \label{eq:sigma_evol}
\end{equation}
which mimics the MHD simulation of \citet{Mertens:2016:AA}.
Outside the fast point, $\sigma$ continuously decreases
and attain $\sigma \ll 1$,
which means that the jet becomes kinetic-dominated asymptotically
\citep[e.g.,][]{Chiueh:1998:ApJ}.
The terminal Lorentz factor tends to $\sigma_0{}^{1/3}$,
where $\sigma_0 (\gg 1)$ denotes the $\sigma$ at the jet launching point.

It is noteworthy that an efficient gamma-ray emission,
as observed during the VHE flares in year 2008 and 2010
\citep{Acciari:2009:Sci,Aliu:2012:ApJ,Abramowski:2012:ApJ},
takes place as a result of synchrotron self-Compton process
only when the jet becomes {\it particle-dominated}
\citep{Blandford:1995:ApJ}.
In addition,
the VHE gamma-rays are appeared to be emitted 
within $20 R_{\rm S}$ from the black hole
\citep{hada13}.
In this context,
the rapid conversion of energy from Poynting to kineitic
within the central mas scales (eq.~[\ref{eq:sigma_evol}])
is consistent with the VHE observations.

\section{Formation of limb-brightened jets} 
\label{sec:jet}
In this section, we examine if a jet appears limb-brightened 
in radio frequencies.
We adopt the photon frequency $\nu= 86$~GHz 
in our frame of reference, and compute expected VLBI maps.

%
%

\subsection{Jet collimation altitude}
\label{sec:jet_collimation}
We follow the method described in \S~\ref{sec:model}
to constrain the emission and absorption coefficients
in the jet co-moving frame (\S~\ref{sec:jet_emissivity}).
Then we apply their Lorentz invariant relations
to convert these quantities in our frame of reference.
Using the obtained emission and absorption coefficients,
we integrate the radiative transfer equation
to compute the specific intensity $I_\nu$ along each line of sight.
Finally, the surface brightness is obtained by multiplying $I_\nu$
with the solid angle subtended by each area on the map.

In this subsection (\S~\ref{sec:jet_collimation}),
we constrain the altitude $r_{\rm c}$ 
below which the jet is collimated,
comparing the results with VLBI observations.
For this purpose, we fix the observer's viewing angle
at $\theta_{\rm v}= 0.30$~rad (i.e., $17^\circ$),
thermal lepton's temperature 
at $\Theta = k T_{\rm e} / m_{\rm e} c^2 = 1.0$,
and the non-thermal lepton's power-law index
at $p= 3.0$.

\begin{figure*}
\vspace*{-5.0 truecm}
\includegraphics[width=\textwidth, angle=0, scale=1.0]{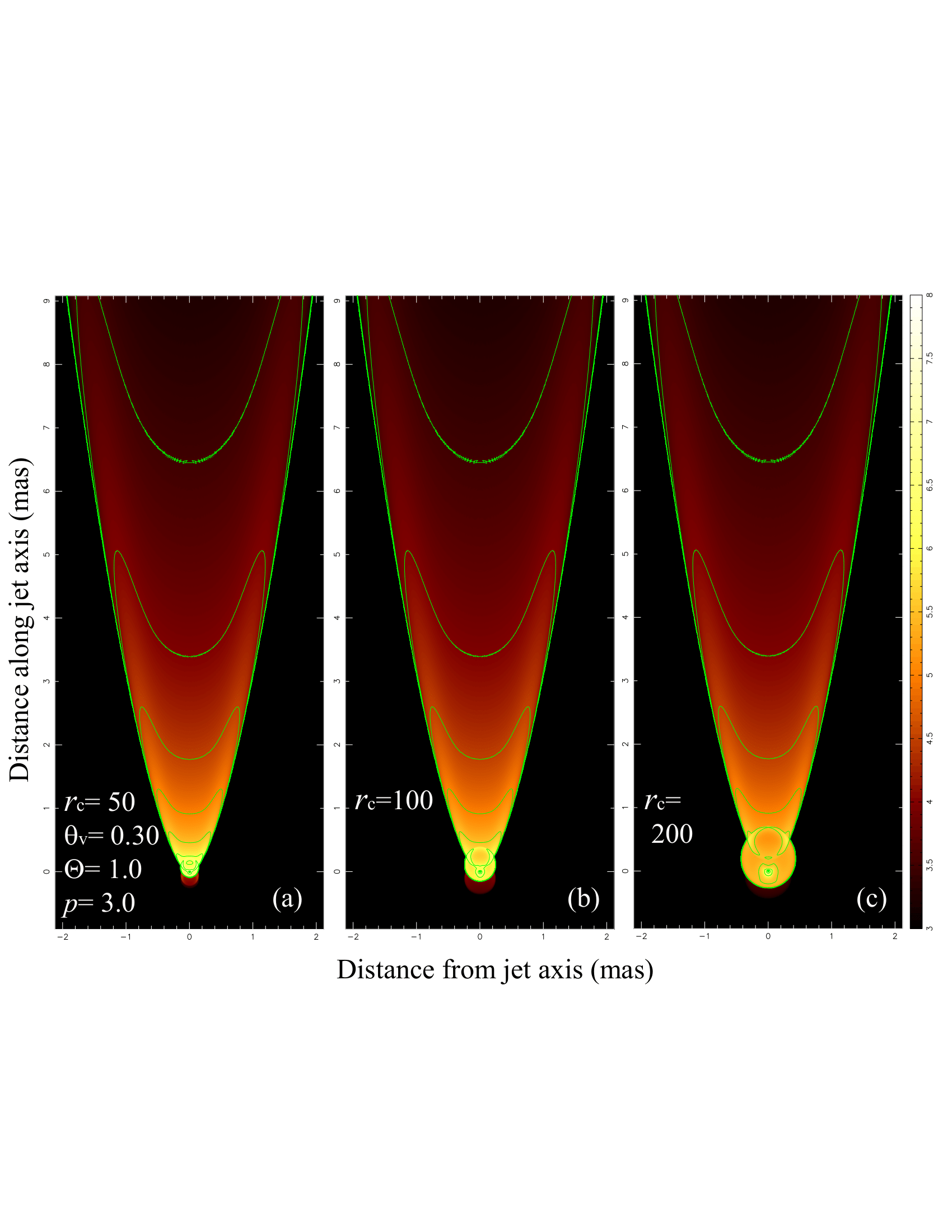}
\vspace*{-5.0 truecm}
\caption{
Surface brightness distribution of synchrotron emission 
from the M87 jet at 86~GHz.
For all the panels, we set
$\theta_{\rm v}= 0.30$~rad,
$\Theta \equiv kT_{\rm e} / m_{\rm e} c^2 = 1.0$, and
$p=3.0$.
In panels (a), (b), and (c), we set
$r_{\rm c}= 50 R_{\rm S}$, $100 R_{\rm S}$, and $200 R_{\rm S}$,
respectively.
Color coding is common to all the panels 
and defined in the color bar on the right.
The contour level increases by $10^{0.5}$ toward the VLBI core, 
which is optically thick for synchrotron self-absorption.
We put $f=1$ in equation~(\ref{eq:Sr_2}) in all panels.
}
    \label{fig:map_collimation}
\end{figure*}

In figure~\ref{fig:map_collimation},
we compare how the jet appears constricted
(perpendicularly to the propagation direction)
when a collimation 
(from conical to parabolic-like flow line)
takes place at various de-projected altitude
$r=r_{\rm c}$ from the BH.
In panels (a), (b), and (c),
we adopt $r_{\rm c}= 50 R_{\rm S}$,
$r_{\rm c}= 100 R_{\rm S}$, and
$r_{\rm c}= 200 R_{\rm S}$, respectively.
It follows that the jet width perpendicular to the 
propagation direction shrinks.
This is because the jet is assumed to change its
shape from conical to parabolic one inside
the de-projected distance $r_{\rm c}$ (eq.~\ref{eq:nu_evolv}),
and because this change of flow-line geometry
leads to a shrink of transverse width in the celestial plane.

The constricted structure appears at
$z= 0.10$~mas, $0.22$~mas, and $0.48$~mas
when $r_{\rm c}/R_{\rm S}= 50$, $100$, and $200$, respectively.
On the other hand, 
VLBI observations of M87 at 86~GHz
reveals that a constricted structure appears
at $z= 0.2$--$0.3$~mas from the VLBI core
\citep{hada16}.
Thus, we consider that the M87 jet finishes its collimation
within a de-projected altitude $r \approx 100 R_{\rm S}$.

\subsection{Limb-brightening of jets}
\label{sec:jet_limb}
Figure~\ref{fig:map_collimation} also shows
that the jet exhibits a limb-brightened structure
for various values of the collimation height $r_{\rm c}$.
This is because the BH's rotational energy is preferentially
extracted along the magnetic field lines threading
the event horizon in the middle or lower latitudes
(fig.~\ref{fig:BZ_theta}).
In another word, the energy carried along each magnetic flux tube
increases with the distance $\varpi$ from the jet axis,
resulting in an efficient mass loading and emission
along the field lines near the jet limb.
In this paper, 
we adjust the normalization constant $A_{\rm max}$
so that the integrated flux density 
of figure~\ref{fig:map_collimation}b
may be consistent with observed values 
($ \sim 1$~Jy at $86$~GHz)
\citep[e.g.,][]{Kim:2018:A&A,hada16}.
We accordingly obtain $B_{\rm p} = 30$~G.

\begin{figure*}
\vspace*{-5.0 truecm}
\includegraphics[width=\textwidth, angle=0, scale=1.0]
{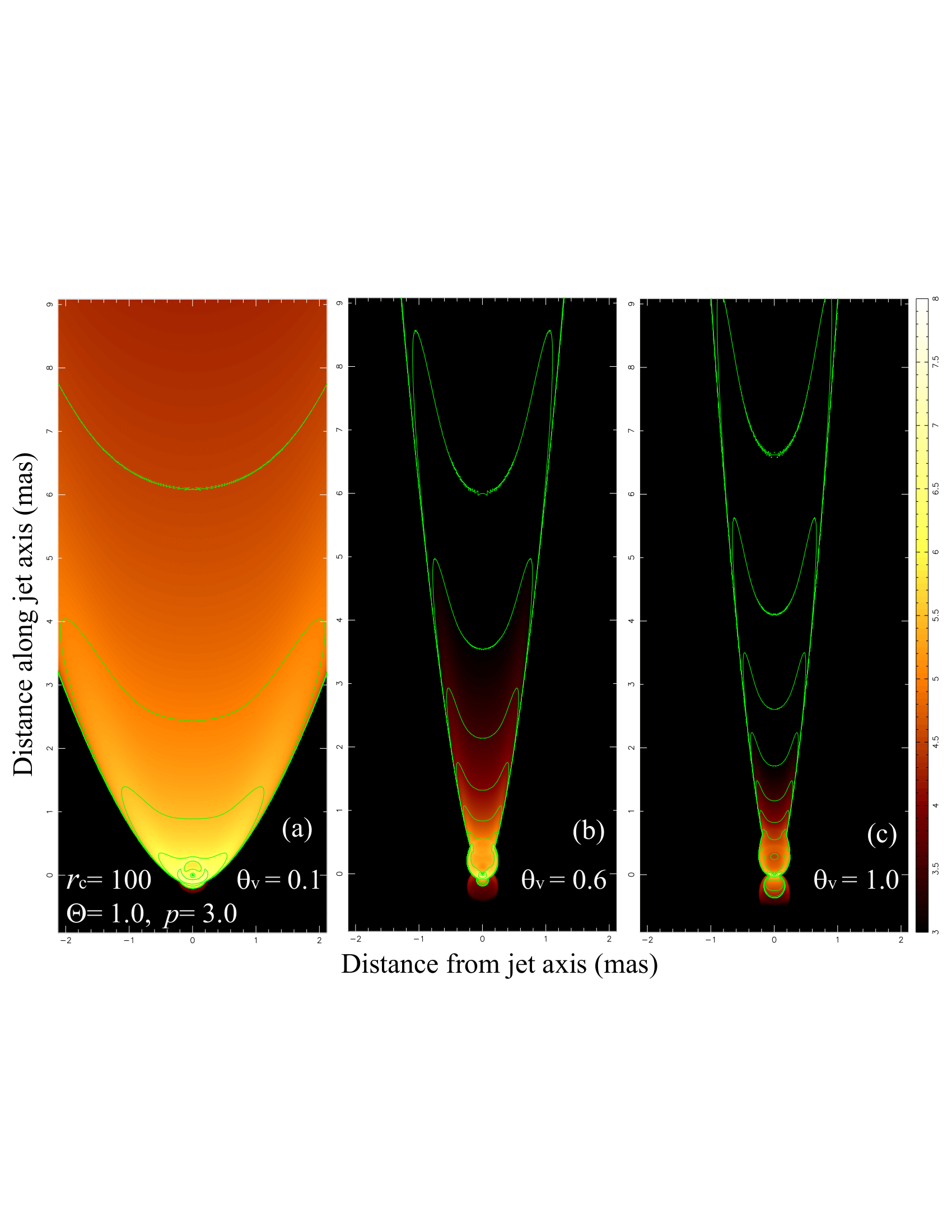}
\vspace*{-5.0 truecm}
\caption{
Similar figure as fig.~\ref{fig:map_collimation},
but the viewing angle is changed.
We set 
$\theta_{\rm v}= 0.1$, $0.6$ and $1.0$ for 
panels (a), (b) and (c), respectively.
Other parameters are commonly set as
$r_{\rm c}= 100 R_{\rm S}$,
$\Theta= 1.0$, and
$p= 3.0$.
Color coding is common to all the panels 
and defined in the color bar on the right.
We put $f=1$ for all panels.
All panels should be compared with fig.~\ref{fig:map_collimation}b.
}
    \label{fig:map_angle}
\end{figure*}

Let us next consider how the limb-brightened jet changes 
its appearance
as a function of the observer's viewing angle, $\theta_{\rm v}$.
In figure~\ref{fig:map_angle}, we adopt 
$\theta_{\rm v}= 0.10$~rad, $0.60$~rad, and $1.00$~rad, 
as denoted in each panel,
keeping other parameter unchanged from
figure~\ref{fig:map_collimation}b.
Comparing figure~\ref{fig:map_angle} with 
figure~\ref{fig:map_collimation}b,
we find that the jet increases its brightness
with decreasing $\theta_{\rm v}$
as a result of the relativistic beaming effect.
It also follows that
the jet perpendicular thickness decreases 
with increasing $\theta_{\rm v}$
as a result of a projection effect.
Namely, the apparent half opening angle,
$\chi_{\rm app}= {\rm atan}(\tan\chi_{\rm int} / \sin\theta_{\rm v})$,
increases with decreasing $\theta_{\rm v}$
for a fixed intrinsic half opening angle, $\chi_{\rm int}$.


It is noteworthy that the limb-brightened structure appears
in a wide range of viewing angle $\theta_{\rm v}$, from
$0.10$~rad (fig.~\ref{fig:map_angle}a), 
$0.30$~rad (fig.~\ref{fig:map_collimation}b), 
$0.60$~rad (fig.~\ref{fig:map_angle}b), to
$1.00$~rad (fig.~\ref{fig:map_angle}c). 
This result is consistent with the VLBI observations,
which find limb-brightened jets in a wide range of $\theta_{\rm v}$.
For example, limb-brightened jets are found
in Mrk~501 for $5^\circ < \theta_{\rm v} < 15^\circ$
\citep{Giroletti:2004:ApJ},
in M87 for $11^\circ < \theta_{\rm v} < 19^\circ$
\citep{Perlman:2011:ApJ, hada16,Kim:2019:A&A},
and in Cyg~A for $\theta_{\rm v} \sim 75^\circ$
\citep{Boccardi:2016:A&A}.


\subsection{Dependence on the magnetic-field geometry}
\label{sec:jet_geometry}
Let us examine how the limb-brightened structure
depends on the large-scale magnetic field configurations.

\subsubsection{Dependence on poloidal field geometry}
\label{sec:jet_geometry_poloidal}
\begin{figure*}
\vspace*{-3.0truecm}
\includegraphics[width=\textwidth, angle=0, scale=1.0]
{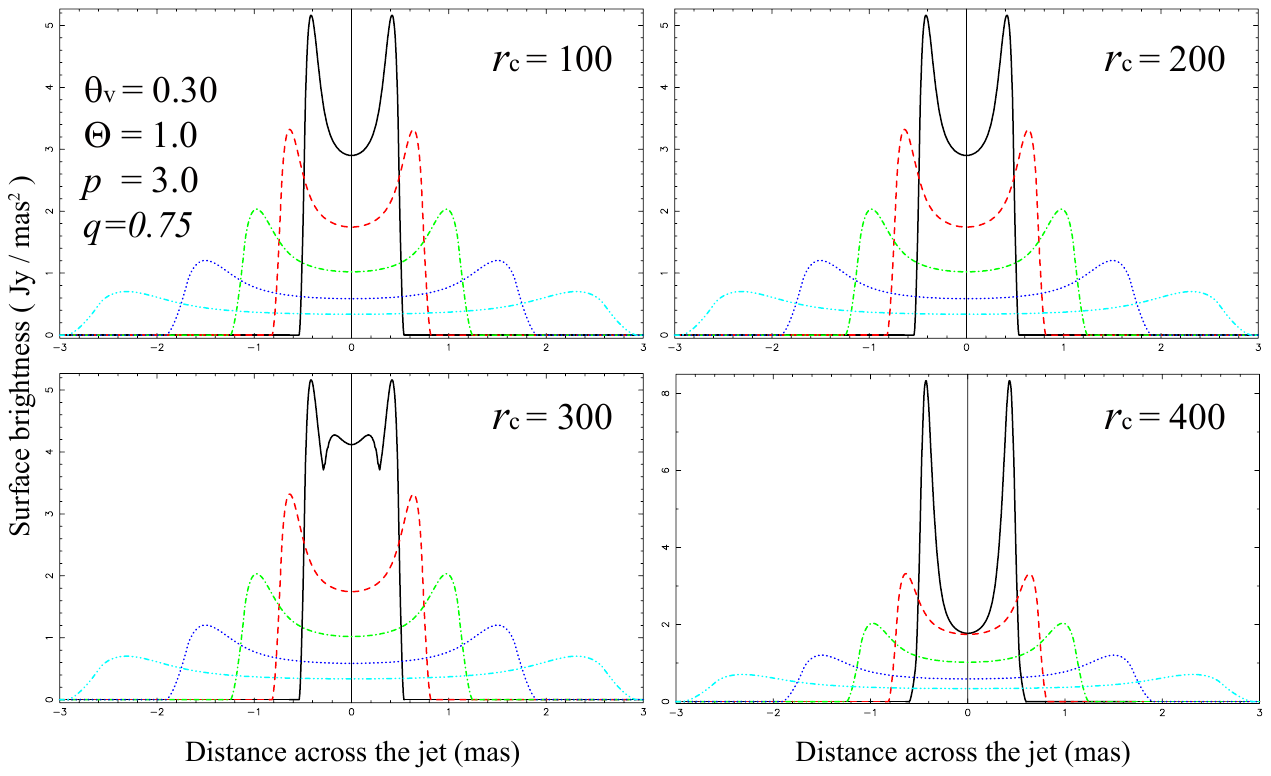}
\vspace*{-8.0truecm}
\caption{
Transverse jet intensity profiles for 
$r_{\rm c}= 100 R_{\rm S}$ (top left),
$200 R_{\rm S}$ (top right),
$300 R_{\rm S}$ (bottom left), and
$400 R_{\rm S}$ (bottom right).
The black solid, red dashed, green dash-dotted, blue dotted, and
cyan dash-dot-dot-dotted curves denote the brightness
at $z= 1$, $2$, $4$, $8$, and $16$~mas from BH.
No difference can be seen between the two top panels.
}
\label{fig:slice_rc}
\end{figure*}

We investigate how the jet structure depends 
on the {\it poloidal} componenents of the magnetic field.
First, we consider its dependence on the collimation altitude, 
$r_{\rm c}$.
In figure~\ref{fig:slice_rc},
we depict the brightness profile 
(or equivalently, the transverse intensity profile)
as a function of $\varpi$ (i.e., the distance across the jet axis),
adopting four discrete collimation altitudes,
$r_{\rm c}= 100 R_{\rm S}$, $200 R_{\rm S}$, $300 R_{\rm S}$, and
$400 R_{\rm S}$.
The black solid, red dashed, green dash-dotted, blue dotted, and
cyan dash-dot-dot-dotted curves show the transverse intensity at
projected distance $z=1.0$~mas, $2.0$~mas, $4.0$~mas, $8.0$~mas, and
$16.0$~mas from the BH.
We fix other parameters than $r_{\rm c}$, adopting
$\theta_{\rm v}= 0.30$~rad,
$\Theta \equiv kT_{\rm e}/m_{\rm e} c^2= 1.0$,
$p=3.0$, and
$q_\infty=0.75$.
It follows from the top two panels of figure~\ref{fig:slice_rc}
that the transverse intensity profile does not change 
as long as $r_{\rm c} < 200 R_{\rm S}$.
From this reason, we constrain $r_{\rm c}$,
using the observed constricted structure of the M87 jet \citep{hada16}, 
instead of the observed transverse intensity profile
\citep[e.g.,][]{Kim:2018:A&A}.
It follows from the bottom two panels of figure~\ref{fig:slice_rc}  
that the limb-brightening is enhanced
when $r_{\rm c} > 300 R_{\rm S}$.
However, if we adopt such a large (de-projected)
collimation altitude,
a constricted structure appears at a much greater
projected distance $z$ than observed \citep{hada16}.
Thus, we do not consider such a large $r_{\rm c}$ 
in what follows.

\begin{figure*}
\vspace*{-5.0truecm}
\includegraphics[width=\textwidth, angle=0, scale=1.0]
{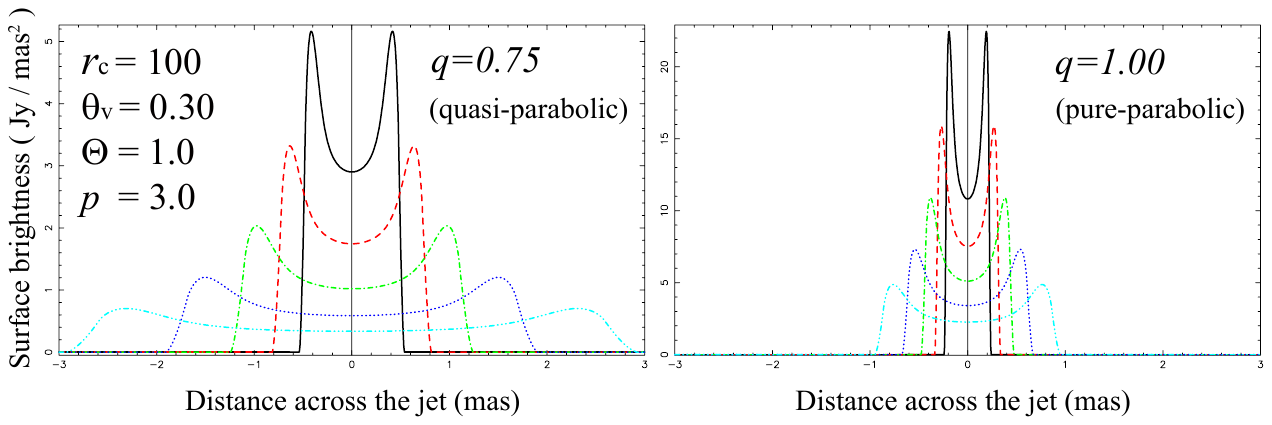}
\vspace*{-8.0truecm}
\caption{
Transverse jet intensity profiles 
for $q_\infty=0.75$ (left) and $q_\infty=1.00$ (right).
The black solid, red dashed, green dash-dotted,
blue dotted, and cyan dash-dot-dot-dotted curves 
show the brightness at altitude
$z= 1$, $2$, $4$, $8$, and $16$~mas 
in the projected distance from the BH along the jet.
The abscissa denotes the distance (mas) measured perpendicular to the jet.
}
\label{fig:slices_q}
\end{figure*}

Second, we consider its dependence on the collimation degree $q$
(eq.~\ref{eq:nu_evolv}).
In figure~\ref{fig:slices_q},
we depict the transverse intensity profile, 
adopting two representative values of $q_\infty$.
The left panel corresponds to the result when the jet geometry
tends to a quasi-parabolic one with $q_\infty= 0.75$
\citep{Asada:2016:ApJ,Nakamura:2018:ApJ}, whereas
the right panel shows the case of a pure-parabolic geometry
(i.e., $q_\infty=1$).
As expected, jet perpendicular width shrinks
when the jet is collimated more strongly
(i.e., when $q$ is greater). 

\begin{figure*}
\vspace*{-5.0truecm}
\includegraphics[width=\textwidth, angle=0, scale=1.0]
{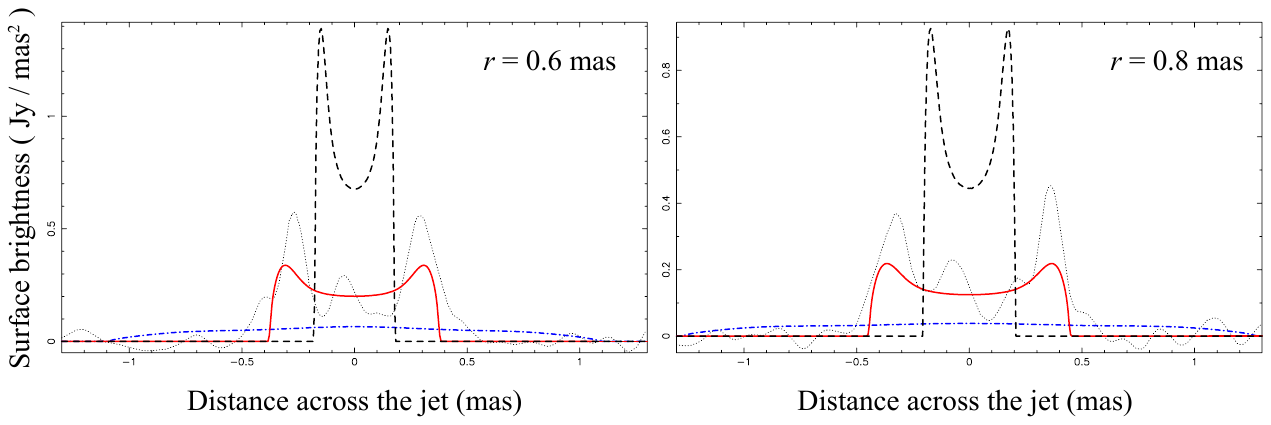}
\vspace*{-8.0truecm}
\caption{
Transverse jet intensity profiles at altitude 
$z=0.6$~mas (left) and $0.8$~mas (right).
The blue dash-dotted, red solid, black dashed curves
denote the brightness computed by the present model for 
$q_\infty=0.50$ (weakly parabolic), 
$q_\infty=0.75$ (quasi parabolic), and 
$q_\infty=1.00$ (purely parabolic), respectively.
The thin dotted curve does 
the observed values reported by 
\citet{Kim:2018:A&A}.
}
\label{fig:slice_rc_obs}
\end{figure*}

It is worth comparing the obtained transverse intensity profile
with observations.
In figure~\ref{fig:slice_rc_obs},
we plot the results at projected distance
$z=0.6$~mas (left) and $z=0.8$~mas (right),
for $q_\infty=0.5$ (blue dash-dotted), 
$q_\infty=0.75$ (red solid), and 
$q_\infty=1.0$ (black dashed).
The thin black dotted curves in each panel denote
what obtained from the M87 jet at $86$~GHz
\citep{Kim:2018:A&A}.
It shows that the limb brightening is more enhanced
when $q_\infty$ increases,
or equivalently, when the jet is more strongly collimated.
We also find that the value $q_\infty=0.75$,
which was inferred by observations at much larger scales,
gives a consistent {\it width} of the brightened limbs.
However, the peak {\it intensity} of the two limbs 
attains only half of what is observed.

\begin{figure*}
\vspace*{-5.0 truecm}
\includegraphics[width=\textwidth, angle=0, scale=1.0]{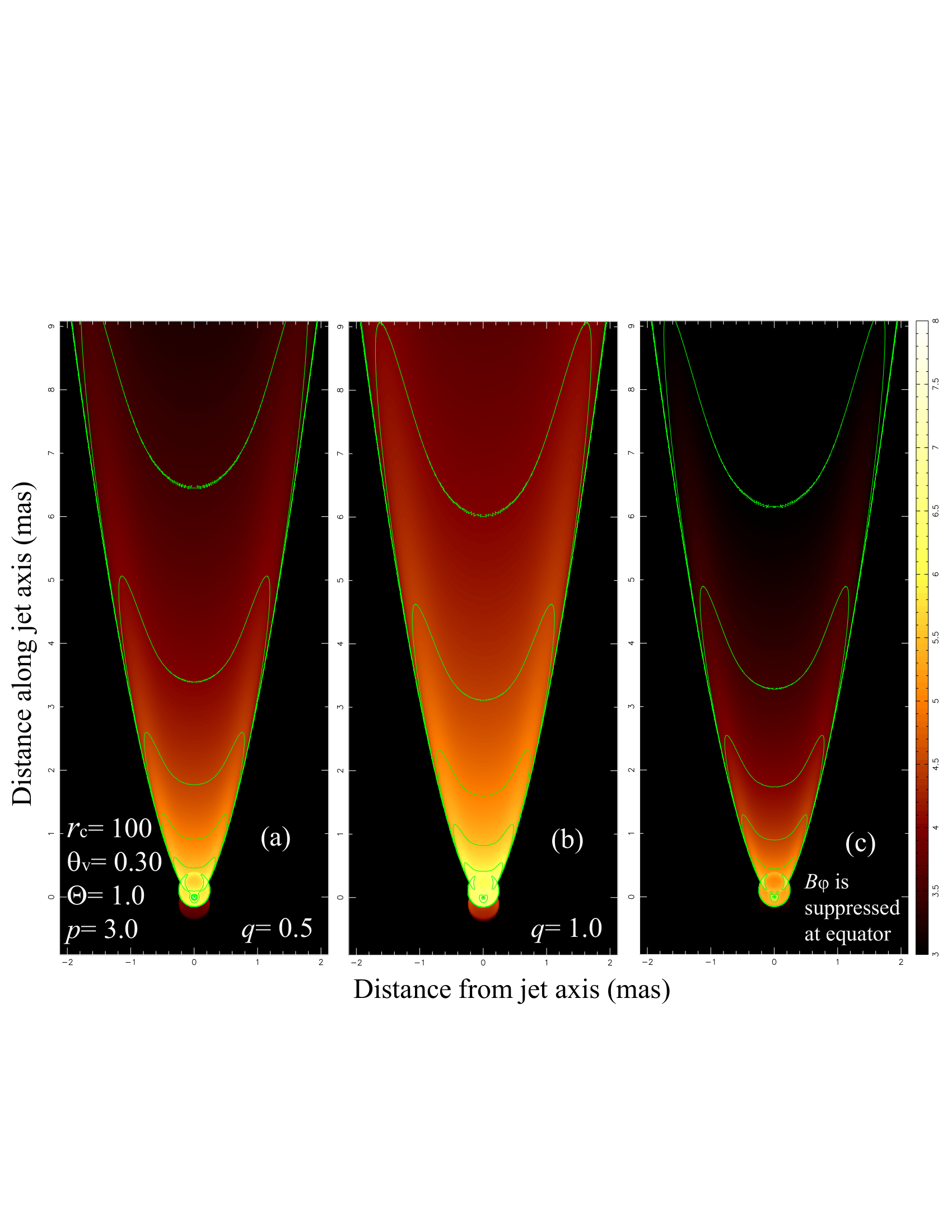}
\vspace*{-5.0 truecm}
\caption{
Similar figure as fig.~\ref{fig:map_collimation},
but the global magnetic field configuration is changed.
{\it Panels (a) and (b):}
We change the poloidal field configuration,
adopting $q_\infty= 0.5$ (weakly parabolic) and 
$q_\infty= 1.0$ (pure parabolic) flow lines in panel (a) and (b),
respectively.
{\it Panels (c):} 
we change the toroidal field configuration,
suppressing $f$ below $1$ near the equator
(black dashed curve in fig.~\ref{fig:BZ_theta}; 
 see also the end of \S~\ref{sec:BZflux}).
All panels should be compared with fig.~\ref{fig:map_collimation}b.
In the present paper,
panel (c) of this figure shows the only case 
in which $f$ deviates from unity in equation~(\ref{eq:Sr_2}). 
}
    \label{fig:map_Bp}
\end{figure*}

We also plot the expected VLBI map in figure~\ref{fig:map_Bp}.
Panels (a) and (b) represent the results for 
$q_\infty=0.5$ and $q_\infty=1.0$,
respectively, and should be compared with the $q_\infty=0.75$ case 
(fig.~\ref{fig:map_collimation}b).
The limb-brightening structure, as well as the constricted structure,
commonly appears for a wide range of $q$.

\subsubsection{Dependence on toroidal field geometry}
\label{sec:jet_geometry_toroidal}
We investigate how the jet structure depends 
on the {\it toroidal} componenents of the magnetic field.
Particularly, we are interested in the case 
when $B_\varphi$, and hence the BZ flux vanishes on the equator.
Thus, to constrain the BZ flux at the jet base,
we adopt the black dashed curve, instead of the red solid one,
in fig.~\ref{fig:BZ_theta}, or equivalently,
we suppress $f(\theta)$ near the equator 
as described at the end of \S~\ref{sec:BZflux}.
We present the resultant VLBI map in figure~\ref{fig:map_Bp}c.
It shows that the jet appearance changes only mildly
when $B_\varphi$ vanishes near the equator.

On these grounds, we consider that main conclusions of the present paper
--- limb brightening and constriction of a jet ---
are robust for a reasonable change of the large-scale magnetic-field geometry.

\subsection{Dependence on lepton energy distribution}
\label{sec:jet_letpon_energy}
\begin{figure*}
\vspace*{-5.0 truecm}
\includegraphics[width=\textwidth, angle=0, scale=1.0]{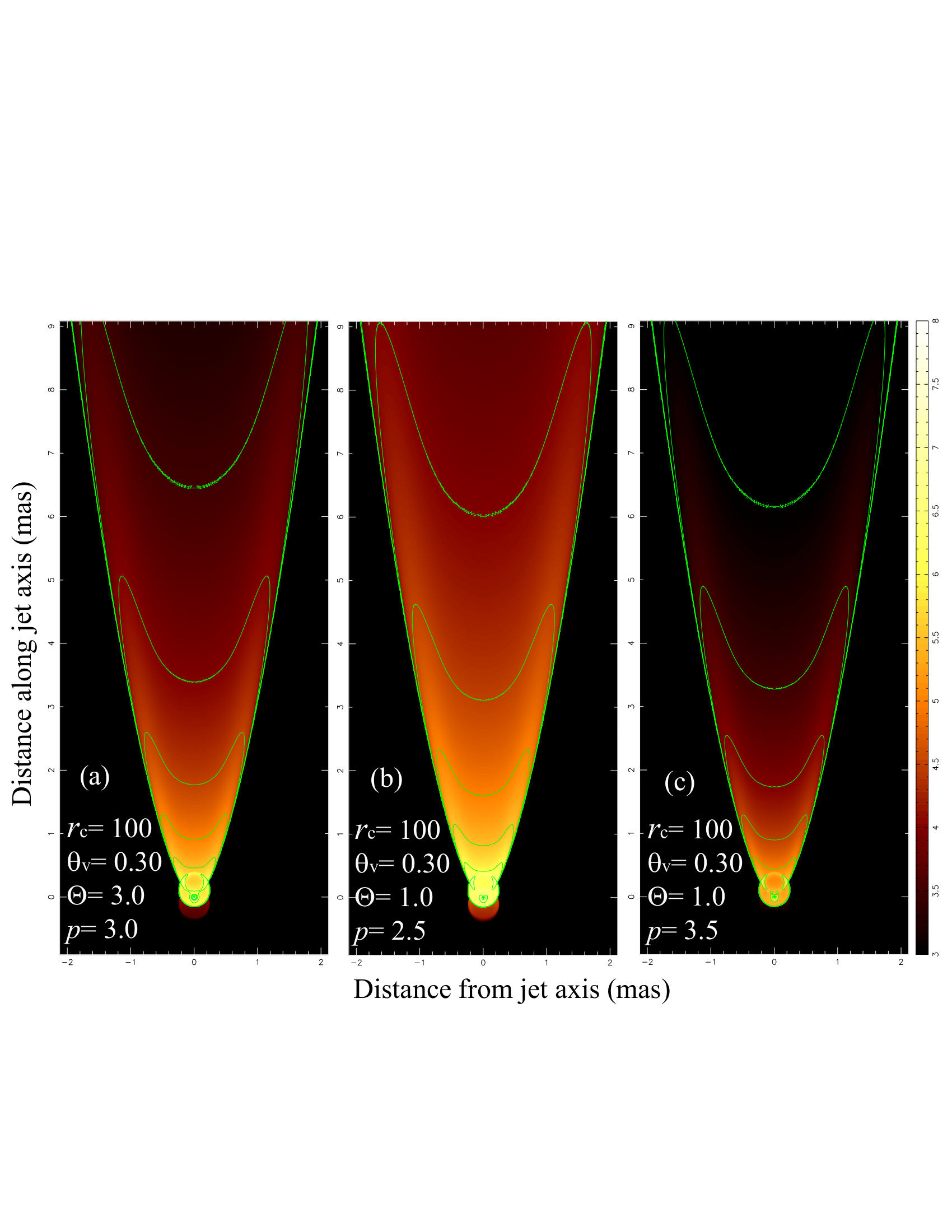}
\vspace*{-5.0 truecm}
\caption{
Similar figure as fig.~\ref{fig:map_collimation},
but the electron energy distribution is changed.
{\it Panel (a):}
we consider higher temperature for thermal leptons,
and put $\Theta= 3.0$, while other parameters are
common with fig~\ref{fig:map_collimation}a.
{\it Panels (b) and (c):}
we consider different power-law index for nonthermal leptons,
and put $p=2.5$ in panel (b) and $p=3.5$ in panel (c).
Color coding is common.
We put $f=1$ in all panels,
which should be compared with fig.~\ref{fig:map_collimation}b.
}
    \label{fig:map_p}
\end{figure*}

We next consider how the surface brightness distribution depends
on the energy distribution of thermal and non-thermal leptons.
In figure~\ref{fig:map_p}a,
we present the result when the thermal leptons
have higher temperature, $k T_{\rm e}= 3.0 m_{\rm e}c^2$.
Comparing with figure~\ref{fig:map_collimation}b,
we find that the limb-brightening feature
is not affected by the lepton temperature.
The difference of temperature appears only at the jet base,
within the central a few mas.

Let us also examine the difference 
when the non-thermal lepton's power-law index $p$ varies.
In figures~\ref{fig:map_p}b and \ref{fig:map_p}c,
we show the cases when $p=2.5$ and $p=3.5$.
Comaparing with figure~\ref{fig:map_collimation}b
(i.e., the case of $p=3.0$),
we find that the limb-brightened structure is not affected
by the changle of $p$, as long as it is between $2.5$ and $3.5$.

\subsection{Counter jet}
\label{sec:jet_counterjet}
The dim structure appearing in $z < 0$ 
in figures~\ref{fig:map_collimation}, \ref{fig:map_angle},
\ref{fig:map_Bp} and \ref{fig:map_p}
shows the counterjet.
The counter jet also exhibits a limb-brightened structure.
Because of the relativistic beaming,
the jet is brighter than the counterjet.
Let us consider the fiducial case of
$r_{\rm c}= 100 R_{\rm S}$,
$\Theta= 1.0$,
$p=3.0$, and
$\theta_{\rm v}= 0.30$~rad
(i.e., fig.~\ref{fig:map_collimation}b).
For the jet, the peak brightness attains
$35.2 \mbox{Jy mas}^{-2}$,
$32.4 \mbox{Jy mas}^{-2}$,
$17.5 \mbox{Jy mas}^{-2}$, and
$9.4 \mbox{Jy mas}^{-2}$ at
$\theta_{\rm v}= 0.10$, $0.30$, $0.60$, and $1.00$, respectively.
For the counterjet, it attains
$0.57 \mbox{Jy mas}^{-2}$,
$0.75 \mbox{Jy mas}^{-2}$,
$0.88 \mbox{Jy mas}^{-2}$,
$1.4 \mbox{Jy mas}^{-2}$, at the same $\theta_{\rm v}$'s.
The integrated flux density within the central $2$ and $5$~mas becomes
$0.11$~Jy and $0.23$~Jy for the jet, whereas it is
$2.2$~mJy and $2.7$~mJy for the coutnerjet.
Accordingly, we find that the jet is brighter than the counterjet
by a factor $R$, 
which lays between $50$ and $85$ in the core region.

\section{Discussion}
\label{sec:disc}
To constrain the angle-dependent energy extraction
from a rotating BH, 
we use the existing MHD simulations of BH magnetospheres
\citep{Tchekhovskoy:2010:ApJ}.
To quantify how such a Poynting flux is converted into 
the jet's kinetic energy flux, 
we use another MHD simulation of 
\citet{Mertens:2016:AA}
for the M87 jet.
Then we compute the lepton density
by the method described in \S~\ref{sec:BZflux},
and demonstrate that the M87 jet naturally exhibits
a limb-brightened structure 
as a result of this angle-dependent energy extraction from the BH.
We also showed that the jet exhibits a constricted structure
because of a collimation from conical to parabolic-like shape.

\subsection{Jet versus counter jet brightness ratio}
\label{sec:disc_ratio}
In \S~\ref{sec:jet_counterjet},
we find that the jet-counterjet brightness ratio becomes
$50 < R < 85$ in the core region.
However, these values are a few times greater than the value
$R=27.1 \pm 9.1$ obtained within the central $0.2$--$0.5$~mas
in the M87 jet \citep[\S~3.6 of][]{Kim:2018:A&A}.
We could in fact, obtain a consistent $R$ with such VLBI observations,
if we adopted a slower evolution of $\Gamma$ or $\sigma$
(than eqs.~\ref{eq:Lf_evol} or \ref{eq:sigma_evol})
as a function of $r$.
However, such a fine-tuning of functions or parameters 
are out of the scope of the present paper,
because the main conclusions 
--- the limb brightening and constriction of a jet ---
are not affected by such details.

\subsection{Formation of kinetic-dominated jets}
\label{sec:disc_cf}
Let us briefly comment on the evolution of the 
magnetization parameter $\sigma$
as described by equation~(\ref{eq:sigma_evol}).
We adopt this equation as an assumption,
because it mimics
the 2D, cold, ideal MHD simulation
of \citet{Mertens:2016:AA},
specifically their models 1 and 2,
which best reproduce the measured acceleration of 
jet components (their \S~4).

However, to the best of the authors' knowledge,
no 3D MHD simulations
that self-consistently solve the system 
from the BH vicinity to the jet downstream,
have successfully demonstrated such a rapid conversion of
the electromagnetic energy into the kinetic one,
e.g., within the initial $100 R_{\rm S}$ 
in the case of the M87 jet.
On the other hand, it is an observational fact
that the jet becomes already kinetic-dominated
\citep{Celotti:1993:MNRAS:264,Blandford:1995:ApJ}
in the high-energy emission region,
which is possibly located within the inner $20 R_{\rm S}$
for the M87 jet 
\citep{hada13}.

To settle down this energy conversion argument,
namely how such a rapid energy conversion could take place
in the jet-launching region,
it may be necessary to quantitatively investigate 
the pair-production cascade in a jet
\citep{Levinson:1995:ApJ,Levinson:1996:ApJL},
and the resultant mass-loading issue.

\subsection{Comparison with Takahashi et al. (2018)}
\label{sec:disc_cf}
To demonstrate the formation of a limb-brightened jet,
we used the angle-dependent energy extraction from the BH,
and constrained the plasma density at each position in the jet.
On the other hand, 
\citet{Takahashi:2018:ApJ}
assumed that the density peaks within a ring-like geometry
away from the jet axis.
In the present paper,
we obtained a symmetric brightness distribution 
with respect to the jet axis, 
because we neglect the rotational motion of fluids 
around the jet axis.
This assumption, a ballistic motion of jet fluids, 
will be justified if the jet is kinetic-dominated.
Since $\sigma \sim 1$ is considered to be achieved 
at the projected distance of $z= 1.5$~mas
\citep{Mertens:2016:AA},
the jet will become kinetic-dominated at $z \gg 2$~mas.
On the other hand, \citet{Takahashi:2018:ApJ} 
took the fluid's rotational motion into account
in the intermediate region where the electromagnetic energy
is being converted into the kinetic energy,
and showed the formation of asymmetric jets 
due to relativistic beaming of radiation.

\subsection{Stronger emissivity along the limb}
\label{sec:disc_limb}
As discussed in \S~\ref{sec:jet_geometry_poloidal},
we failed to obtain a strong enough intensity at the limbs,
comparing with what are obtained in the M87 jet
(fig.~\ref{fig:slice_rc_obs}).
This might be because we adopted a stationary, ideal MHD model
in the present paper.
Nevertheless, it is likely that the jet carrying strong energy flux
near the limb (fig.~\ref{fig:BZ_theta}) 
causes fluid instability or magnetic reconnection at the layer 
with the disk wind, and accelerate nonthermal particles 
at the jet limb
\citep{Matsumoto:2013:ApJL,Parfrey:2015:MNRAS}.
Alternatively, pair-production cascade taking place
in a BH magnetosphere
\citep{bes92,Hirotani:1998:ApJ,nero07,Rieger:2008:A&A,levi11,Broderick:2015:ApJ,hiro16a}
could result in a angle-dependent mass loading
in the jet launching region.
This effect works most efficiently
when the mass accretion rate is so small 
that the gap longtitudinal (radial) thickness becomes
comparable to the transverse width in the meridional direction
\citep{Hirotani:2016:ApJ}.
In this case, the voltage drop along magnetic field lines
becomes a good fraction of the electro-motive force exerted
on the BH surface.
Accordingly, the sum of the kinetic energy density 
of the accelerated charges
and the radiation energy density of the emitted photons
becomes a good fraction the typical electromagnetic energy density
near the BH.
If it happens along the magnetic field lines threading horizon
in the middle latitude,
stronger emission may be expected along such flow lines,
resulting in a stronger emission from the limbs.

\subsection{Formation of the central-ridge emission}
\label{sec:disc_ridge}
VLBI observations have revealed that the M87 jet has a triple-ridge structure
from the jet base whose distance from the central ring structure is
approximately $30 R_{\rm S}$ 
\citep{Lu:2023:Natur}
to larger scales whose distance is greater than $100 R_{\rm S}$
\citep{Kim:2018:A&A}.
Near the jet base, the intensity of the the central-ridge structure 
(near the jet axis) attains about $60$~\% of that of 
the outer-ridge (i.e., the limb).
At larger scales, it attains about $45$~\% of that of the outer ridge.

On these grounds, as long as the M87 jet is concerned,
it is not enough to discuss only the two brightened limbs.
Nevertheless, we consider that the central ridge emission
is formed when the polar funnel of the BH magnetosphere
is failed to be described within the MHD framework,
which we have based on in the present paper.
For instance, a strong BH gap is formed in the polar region
when the accretion rate is much small compared to the Eddignton rate,
and when the BH is extremely rotating
\citep{Song:2017:MNRAS}.
Therefore, taking account of such non-force-free or non-MHD effects, 
we will investigate if the emission is enhanced
along the jet axis in our subsequent papers.

\begin{acknowledgments}
The authors would like to thank the anonymous referee whose comments and suggestions have significantly improved the presentation of this work. The authors acknowledge grant support for the CompAS group under Theory from the Institute of Astronomy and Astrophysics, Academia Sinica (ASIAA), and the National Science and Technology Council (NSTC) in Taiwan through grants 110-2112-M-001-019-, 111-2112-M-001-074-, and 112-2112-M-001-030-. The authors acknowledge the access to high-performance facilities (TIARA cluster and storage) in ASIAA and thank the National Center for High-performance Computing (NCHC) of National Applied Research Laboratories (NARLabs) in Taiwan for providing computational and storage resources. This work utilized tools (BIWA GRPIC code) developed and maintained by the ASIAA CompAS group.
This research has made use of SAO/NASA Astrophysics Data System.

\end{acknowledgments}

\appendix

\section{Emission and absorption coefficients}
\label{sec:app_coeff}
In this appendix, 
we present the expressions of the emission and absorption coefficeints
for thermal and non-thermal electron-positron pairs.
All the quantities are expressed in the jet co-moving frame.

For thermal pairs, we assume that their energy distribution obeys 
the Maxwell-J$\ddot{\rm u}$ttner distribution.
We then obtain the emission coefficient
\citep{Leung:2011:ApJ,Wardzinski:2000:MNRAS},
\begin{equation}
  j_\nu^{\rm (th)}
  = \frac{2\sqrt{2}\pi}{27}
    \frac{\Theta^2}{K_2(\Theta^{-1})}
    \frac{e^2}{c}
    \nu_{\rm L}
    n_\ast
    X^{1/3} \left( X^{1/3} + 2^{11/12} \right)^2
    \exp\left( -X^{1/3} \right),
    \label{eq:emis_th}
\end{equation}
where $X \equiv \nu / \nu_{\rm s}$,
$\nu_{\rm s} \equiv (2/9) \nu_{\rm L} \Theta^2 \sin\chi$,
$\nu_{\rm L} \equiv eB \sin\chi / ( 2\pi m_{\rm e} c)
 = 2.799 \times 10^6 B \sin\chi$;
$K_2$ denotes the modified Bessel function of the second kind 
of order $2$.
The pitch angle can be set as $\chi \sim 60^\circ$.
In this appendix, $B$ denotes the magnetic field strength
and $\nu$ does the photon frequency,
both in the jet co-moving frame.

The absorption coefficient can be obtained by applying 
the Kirchhoff's law,
\begin{equation}
  \alpha_\nu^{\rm (th)}
  = \frac{j_\nu^{\rm (th)}}{B_\nu(T_{\rm e})},
  \label{eq:abso_th}
\end{equation}
where $B_\nu(T_{\rm e})$ denotes the Planck function.
At radio frequencies, we can put
$h\nu \ll \Theta m_{\rm e}c^2 = k T_{\rm e}$
and apply the Reighley-Jeans law,
$B_\nu(T_{\rm e}) \approx 2 m_{\rm e} \nu^2 \Theta$.

For non-thermal pairs, we assume a power law energy distribution,
\begin{equation}
  \frac{d n_\ast}{d\gamma}
  = n_0 \gamma^{-p},
\end{equation}
where $\gamma$ denotes the Lorentz factor associated
with their random motion, and $p$ the power-law index.
Assuming that this power-law holds
within the Lorentz factor range,
$\gamma_{\rm min} < \gamma < \gamma_{\rm max}$,
we obtain
\begin{equation}
  n_\ast
  = \int_{\gamma_{\rm min}}^{\gamma_{\rm max}} 
        \frac{dn_\ast}{d\gamma} d\gamma
  = \frac{\gamma_{\rm min}^{1-p}-\gamma_{\rm max}^{1-p}}
           {p-1}
      n_0
\end{equation}
Assuming 
$\nu \gg (3/2) \gamma_{\rm min}^2 \nu_{\rm L}$ and
$\nu \ll (3/2) \gamma_{\rm max}^2 \nu_{\rm L}$,
we obtain the emission coefficient
\citep{Rybicki:1986:rpa},
\begin{equation}
  \j_\nu^{\rm (nt)}
  = \frac{\sqrt{3}}{2} \frac{e^2}{c} \nu_{\rm L} n_0
    \frac{\displaystyle
          \Gamma \left( \frac{p}{4}+\frac{19}{12} \right)
          \Gamma \left( \frac{p}{4}-\frac{ 1}{12} \right)}{p+1}
    \left( \frac13 \frac{\nu}{\nu_{\rm L}} \right)^{-(p-1)/2}
  \label{eq:emis_nt}
\end{equation}
The absorption coefficient becomes
\citep{LeRoux:1961,Ginzburg:1965}
\begin{equation}
  \alpha_\nu^{\rm (nt)}
  = C(\alpha) \sin\chi \cdot n_0 r_0{}^2 \frac{\nu_0}{\nu_{\rm L}}
    \left( \frac{1}{\sin\chi} \frac{\nu_{\rm L}}{\nu} \right)^{2+p/2}.
  \label{eq:abso_nt}
\end{equation}

\bibliographystyle{aasjournal}

\begin{thebibliography}{}
\expandafter\ifx\csname natexlab\endcsname\relax\def\natexlab#1{#1}\fi
\providecommand{\url}[1]{\href{#1}{#1}}
\providecommand{\dodoi}[1]{doi:~\href{http://doi.org/#1}{\nolinkurl{#1}}}
\providecommand{\doeprint}[1]{\href{http://ascl.net/#1}{\nolinkurl{http://ascl.net/#1}}}
\providecommand{\doarXiv}[1]{\href{https://arxiv.org/abs/#1}{\nolinkurl{https://arxiv.org/abs/#1}}}

\bibitem[{{Abramowski} {et~al.}(2012){Abramowski}, {Acero}, {Aharonian},
  {Akhperjanian}, {Anton}, {Balzer}, {Barnacka}, {Barres de Almeida},
  {Becherini}, {Becker}, {Behera}, {Bernl{\"o}hr}, {Birsin}, {Biteau},
  {Bochow}, {Boisson}, {Bolmont}, {Bordas}, {Brucker}, {Brun}, {Brun}, {Bulik},
  {B{\"u}sching}, {Carrigan}, {Casanova}, {Cerruti}, {Chadwick}, {Charbonnier},
  {Chaves}, {Cheesebrough}, {Clapson}, {Coignet}, {Cologna}, {Conrad},
  {Dalton}, {Daniel}, {Davids}, {Degrange}, {Deil}, {Dickinson},
  {Djannati-Ata{\"\i}}, {Domainko}, {Drury}, {Dubus}, {Dutson}, {Dyks},
  {Dyrda}, {Egberts}, {Eger}, {Espigat}, {Fallon}, {Farnier}, {Fegan},
  {Feinstein}, {Fernandes}, {Fiasson}, {Fontaine}, {F{\"o}rster},
  {F{\"u}{\ss}ling}, {Gallant}, {Gast}, {G{\'e}rard}, {Gerbig}, {Giebels},
  {Glicenstein}, {Gl{\"u}ck}, {Goret}, {G{\"o}ring}, {H{\"a}ffner}, {Hague},
  {Hampf}, {Hauser}, {Heinz}, {Heinzelmann}, {Henri}, {Hermann}, {Hinton},
  {Hoffmann}, {Hofmann}, {Hofverberg}, {Holler}, {Horns}, {Jacholkowska}, {de
  Jager}, {Jahn}, {Jamrozy}, {Jung}, {Kastendieck}, {Katarzy{\'n}ski}, {Katz},
  {Kaufmann}, {Keogh}, {Khangulyan}, {Kh{\'e}lifi}, {Klochkov}, {Klu{\'z}niak},
  {Kneiske}, {Komin}, {Kosack}, {Kossakowski}, {Laffon}, {Lamanna}, {Lennarz},
  {Lohse}, {Lopatin}, {Lu}, {Marandon}, {Marcowith}, {Masbou}, {Maurin},
  {Maxted}, {Mayer}, {McComb}, {Medina}, {M{\'e}hault}, {Moderski}, {Moulin},
  {Naumann}, {Naumann-Godo}, {de Naurois}, {Nedbal}, {Nekrassov}, {Nguyen},
  {Nicholas}, {Niemiec}, {Nolan}, {Ohm}, {de O{\~n}a Wilhelmi}, {Opitz},
  {Ostrowski}, {Oya}, {Panter}, {Paz Arribas}, {Pedaletti}, {Pelletier},
  {Petrucci}, {Pita}, {P{\"u}hlhofer}, {Punch}, {Quirrenbach}, {Raue},
  {Rayner}, {Reimer}, {Reimer}, {Renaud}, {de los Reyes}, {Rieger}, {Ripken},
  {Rob}, {Rosier-Lees}, {Rowell}, {Rudak}, {Rulten}, {Ruppel}, {Sahakian},
  {Sanchez}, {Santangelo}, {Schlickeiser}, {Sch{\"o}ck}, {Schulz}, {Schwanke},
  {Schwarzburg}, {Schwemmer}, {Sheidaei}, {Skilton}, {Sol}, {Spengler},
  {Stawarz}, {Steenkamp}, {Stegmann}, {Stinzing}, {Stycz}, {Sushch}, {Szostek},
  {Tavernet}, {Terrier}, {Tluczykont}, {Valerius}, {van Eldik}, {Vasileiadis},
  {Venter}, {Vialle}, {Viana}, {Vincent}, {V{\"o}lk}, {Volpe}, {Vorobiov},
  {Vorster}, {Wagner}, {Ward}, {White}, {Wierzcholska}, {Zacharias}, {Zajczyk},
  {Zdziarski}, {Zech}, {Zechlin}, {H.~E.~S.~S. Collaboration}, {Aleksi{\'c}},
  {Antonelli}, {Antoranz}, {Backes}, {Barrio}, {Bastieri}, {Becerra
  Gonz{\'a}lez}, {Bednarek}, {Berdyugin}, {Berger}, {Bernardini}, {Biland},
  {Blanch}, {Bock}, {Boller}, {Bonnoli}, {Borla Tridon}, {Braun}, {Bretz},
  {Ca{\~n}ellas}, {Carmona}, {Carosi}, {Colin}, {Colombo}, {Contreras},
  {Cortina}, {Cossio}, {Covino}, {Dazzi}, {De Angelis}, {De Cea del Pozo}, {De
  Lotto}, {Delgado Mendez}, {Diago Ortega}, {Doert}, {Dom{\'\i}nguez}, {Dominis
  Prester}, {Dorner}, {Doro}, {Elsaesser}, {Ferenc}, {Fonseca}, {Font},
  {Fruck}, {Garc{\'\i}a L{\'o}pez}, {Garczarczyk}, {Garrido}, {Giavitto},
  {Godinovi{\'c}}, {Hadasch}, {H{\"a}fner}, {Herrero}, {Hildebrand},
  {H{\"o}hne-M{\"o}nch}, {Hose}, {Hrupec}, {Huber}, {Jogler}, {Klepser},
  {Kr{\"a}henb{\"u}hl}, {Krause}, {La Barbera}, {Lelas}, {Leonardo},
  {Lindfors}, {Lombardi}, {L{\'o}pez}, {Lorenz}, {Makariev}, {Maneva},
  {Mankuzhiyil}, {Mannheim}, {Maraschi}, {Mariotti}, {Mart{\'\i}nez}, {Mazin},
  {Meucci}, {Miranda}, {Mirzoyan}, {Miyamoto}, {Mold{\'o}n}, {Moralejo},
  {Munar}, {Nieto}, {Nilsson}, {Orito}, {Oya}, {Paneque}, {Paoletti}, {Pardo},
  {Paredes}, {Partini}, {Pasanen}, {Pauss}, {Perez-Torres}, {Persic},
  {Peruzzo}, {Pilia}, {Pochon}, {Prada}, {Prada Moroni}, {Prandini}, {Puljak},
  {Reichardt}, {Reinthal}, {Rhode}, {Rib{\'o}}, {Rico}, {R{\"u}gamer},
  {Saggion}, {Saito}, {Saito}, {Salvati}, {Satalecka}, {Scalzotto}, {Scapin},
  {Schultz}, {Schweizer}, {Shayduk}, {Shore}, {Sillanp{\"a}{\"a}}, {Sitarek},
  {Sobczynska}, {Spanier}, {Spiro}, {Stamerra}, {Steinke}, {Storz}, {Strah},
  {Suri{\'c}}, {Takalo}, {Takami}, {Tavecchio}, {Temnikov}, {Terzi{\'c}},
  {Tescaro}, {Teshima}, {Thom}, {Tibolla}, {Torres}, {Treves}, {Vankov},
  {Vogler}, {Wagner}, {Weitzel}, {Zabalza}, {Zandanel}, {Zanin}, {MAGIC
  Collaboration}, {Arlen}, {Aune}, {Beilicke}, {Benbow}, {Bouvier}, {Bradbury},
  {Buckley}, {Bugaev}, {Byrum}, {Cannon}, {Cesarini}, {Ciupik}, {Connolly},
  {Cui}, {Dickherber}, {Duke}, {Errando}, {Falcone}, {Finley}, {Finnegan},
  {Fortson}, {Furniss}, {Galante}, {Gall}, {Godambe}, {Griffin}, {Grube},
  {Gyuk}, {Hanna}, {Holder}, {Huan}, {Hui}, {Kaaret}, {Karlsson}, {Kertzman},
  {Khassen}, {Kieda}, {Krawczynski}, {Krennrich}, {Lang}, {LeBohec}, {Maier},
  {McArthur}, {McCann}, {Moriarty}, {Mukherjee}, {Nu{\~n}ez}, {Ong}, {Orr},
  {Otte}, {Park}, {Perkins}, {Pichel}, {Pohl}, {Prokoph}, {Ragan}, {Reyes},
  {Reynolds}, {Roache}, {Rose}, {Ruppel}, {Schroedter}, {Sembroski},
  {{\c{S}}ent{\"u}rk}, {Telezhinsky}, {Te{\v{s}}i{\'c}}, {Theiling},
  {Thibadeau}, {Varlotta}, {Vassiliev}, {Vivier}, {Wakely}, {Weekes},
  {Williams}, {Zitzer}, {VERITAS Collaboration}, {Barres de Almeida}, {Cara},
  {Casadio}, {Cheung}, {McConville}, {Davies}, {Doi}, {Giovannini},
  {Giroletti}, {Hada}, {Hardee}, {Harris}, {Junor}, {Kino}, {Lee}, {Ly},
  {Madrid}, {Massaro}, {Mundell}, {Nagai}, {Perlman}, {Steele}, {Walker}, \&
  {Wood}}]{Abramowski:2012:ApJ}
{Abramowski}, A., {Acero}, F., {Aharonian}, F., {et~al.} 2012, \apj, 746, 151,
  \dodoi{10.1088/0004-637X/746/2/151}

\bibitem[{{Acciari} {et~al.}(2009){Acciari}, {Aliu}, {Arlen}, {Bautista},
  {Beilicke}, {Benbow}, {Bradbury}, {Buckley}, {Bugaev}, {Butt}, {Byrum},
  {Cannon}, {Celik}, {Cesarini}, {Chow}, {Ciupik}, {Cogan}, {Cui},
  {Dickherber}, {Fegan}, {Finley}, {Fortin}, {Fortson}, {Furniss}, {Gall},
  {Gillanders}, {Grube}, {Guenette}, {Gyuk}, {Hanna}, {Holder}, {Horan}, {Hui},
  {Humensky}, {Imran}, {Kaaret}, {Karlsson}, {Kieda}, {Kildea}, {Konopelko},
  {Krawczynski}, {Krennrich}, {Lang}, {LeBohec}, {Maier}, {McCann},
  {McCutcheon}, {Millis}, {Moriarty}, {Ong}, {Otte}, {Pandel}, {Perkins},
  {Petry}, {Pohl}, {Quinn}, {Ragan}, {Reyes}, {Reynolds}, {Roache}, {Roache},
  {Rose}, {Schroedter}, {Sembroski}, {Smith}, {Swordy}, {Theiling}, {Toner},
  {Varlotta}, {Vincent}, {Wakely}, {Ward}, {Weekes}, {Weinstein}, {Williams},
  {Wissel}, {Wood}, {Walker}, {Davies}, {Hardee}, {Junor}, {Ly}, {Aharonian},
  {Akhperjanian}, {Anton}, {Barres de Almeida}, {Bazer-Bachi}, {Becherini},
  {Behera}, {Bernl{\"o}hr}, {Bochow}, {Boisson}, {Bolmont}, {Borrel},
  {Brucker}, {Brun}, {Brun}, {B{\"u}hler}, {Bulik}, {B{\"u}sching},
  {Boutelier}, {Chadwick}, {Charbonnier}, {Chaves}, {Cheesebrough}, {Chounet},
  {Clapson}, {Coignet}, {Dalton}, {Daniel}, {Davids}, {Degrange}, {Deil},
  {Dickinson}, {Djannati-Ata{\"\i}}, {Domainko}, {Drury}, {Dubois}, {Dubus},
  {Dyks}, {Dyrda}, {Egberts}, {Emmanoulopoulos}, {Espigat}, {Farnier},
  {Feinstein}, {Fiasson}, {F{\"o}rster}, {Fontaine}, {F{\"u}{\ss}ling},
  {Gabici}, {Gallant}, {G{\'e}rard}, {Gerbig}, {Giebels}, {Glicenstein},
  {Gl{\"u}ck}, {Goret}, {G{\"o}hring}, {Hauser}, {Hauser}, {Heinz},
  {Heinzelmann}, {Henri}, {Hermann}, {Hinton}, {Hoffmann}, {Hofmann},
  {Holleran}, {Hoppe}, {Horns}, {Jacholkowska}, {de Jager}, {Jahn}, {Jung},
  {Katarzy{\'n}ski}, {Katz}, {Kaufmann}, {Kendziorra}, {Kerschhaggl},
  {Khangulyan}, {Kh{\'e}lifi}, {Keogh}, {Klu{\'z}niak}, {Kneiske}, {Komin},
  {Kosack}, {Lamanna}, {Lenain}, {Lohse}, {Marandon}, {Martin},
  {Martineau-Huynh}, {Marcowith}, {Maurin}, {McComb}, {Medina}, {Moderski},
  {Moulin}, {Naumann-Godo}, {de Naurois}, {Nedbal}, {Nekrassov}, {Nicholas},
  {Niemiec}, {Nolan}, {Ohm}, {Olive}, {O{\~n}a de Wilhelmi}, {Orford},
  {Ostrowski}, {Panter}, {Paz Arribas}, {Pedaletti}, {Pelletier}, {Petrucci},
  {Pita}, {P{\"u}hlhofer}, {Punch}, {Quirrenbach}, {Raubenheimer}, {Raue},
  {Rayner}, {Renaud}, {Rieger}, {Ripken}, {Rob}, {Rosier-Lees}, {Rowell},
  {Rudak}, {Rulten}, {Ruppel}, {Sahakian}, {Santangelo}, {Schlickeiser},
  {Sch{\"o}ck}, {Schr{\"o}der}, {Schwanke}, {Schwarzburg}, {Schwemmer},
  {Shalchi}, {Sikora}, {Skilton}, {Sol}, {Spangler}, {Stawarz}, {Steenkamp},
  {Stegmann}, {Stinzing}, {Superina}, {Szostek}, {Tam}, {Tavernet}, {Terrier},
  {Tibolla}, {Tluczykont}, {van Eldik}, {Vasileiadis}, {Venter}, {Venter},
  {Vialle}, {Vincent}, {Vivier}, {V{\"o}lk}, {Volpe}, {Wagner}, {Ward},
  {Zdziarski}, {Zech}, {Anderhub}, {Antonelli}, {Antoranz}, {Backes},
  {Baixeras}, {Balestra}, {Barrio}, {Bastieri}, {Becerra Gonz{\'a}lez},
  {Becker}, {Bednarek}, {Berger}, {Bernardini}, {Biland}, {Bock}, {Bonnoli},
  {Bordas}, {Tridon}, {Bosch-Ramon}, {Bose}, {Braun}, {Bretz}, {Britvitch},
  {Camara}, {Carmona}, {Commichau}, {Contreras}, {Cortina}, {Costado},
  {Covino}, {Curtef}, {Dazzi}, {De Angelis}, {de Cea del Pozo}, {Delgado
  Mendez}, {De los Reyes}, {De Lotto}, {De Maria}, {De Sabata}, {Dominguez},
  {Dorner}, {Doro}, {Elsaesser}, {Errando}, {Ferenc}, {Fern{\'a}ndez}, {Firpo},
  {Fonseca}, {Font}, {Galante}, {Garc{\'\i}a L{\'o}pez}, {Garczarczyk}, {Gaug},
  {Goebel}, {Hadasch}, {Hayashida}, {Herrero}, {Hildebrand},
  {H{\"o}hne-M{\"o}nch}, {Hose}, {Hsu}, {Jogler}, {Kranich}, {La Barbera},
  {Laille}, {Leonardo}, {Lindfors}, {Lombardi}, {Longo}, {L{\'o}pez}, {Lorenz},
  {Majumdar}, {Maneva}, {Mankuzhiyil}, {Mannheim}, {Maraschi}, {Mariotti},
  {Mart{\'\i}nez}, {Mazin}, {Meucci}, {Miranda}, {Mirzoyan}, {Miyamoto},
  {Mold{\'o}n}, {Moles}, {Moralejo}, {Nieto}, {Nilsson}, {Ninkovic}, {Oya},
  {Paoletti}, {Paredes}, {Pasanen}, {Pascoli}, {Pauss}, {Pegna},
  {Perez-Torres}, {Persic}, {Peruzzo}, {Prada}, {Prandini}, {Puchades},
  {Reichardt}, {Rhode}, {Rib{\'o}}, {Rico}, {Rissi}, {Robert}, {R{\"u}gamer},
  {Saggion}, {Saito}, {Salvati}, {Sanchez-Conde}, {Satalecka}, {Scalzotto},
  {Scapin}, {Schweizer}, {Shayduk}, {Shore}, {Sidro}, {Sierpowska-Bartosik},
  {Sillanp{\"a}{\"a}}, {Sitarek}, {Sobczynska}, {Spanier}, {Stamerra}, {Stark},
  {Takalo}, {Tavecchio}, {Temnikov}, {Tescaro}, {Teshima}, {Torres}, {Turini},
  {Vankov}, {Wagner}, {Zabalza}, {Zandanel}, {Zanin}, {Zapatero}, {VERITAS
  Collaboration}, {VLBA 43 GHz M87 Monitoring Team}, {H.~E.~S.~S.
  Collaboration}, \& {MAGIC Collaboration}}]{Acciari:2009:Sci}
{Acciari}, V.~A., {Aliu}, E., {Arlen}, T., {et~al.} 2009, Science, 325, 444,
  \dodoi{10.1126/science.1175406}

\bibitem[{{Aliu} {et~al.}(2012){Aliu}, {Arlen}, {Aune}, {Beilicke}, {Benbow},
  {Bouvier}, {Bradbury}, {Buckley}, {Bugaev}, {Byrum}, {Cannon}, {Cesarini},
  {Ciupik}, {Collins-Hughes}, {Connolly}, {Cui}, {Dickherber}, {Duke},
  {Errando}, {Falcone}, {Finley}, {Finnegan}, {Fortson}, {Furniss}, {Galante},
  {Gall}, {Godambe}, {Griffin}, {Grube}, {Guenette}, {Gyuk}, {Hanna}, {Holder},
  {Huan}, {Hughes}, {Hui}, {Humensky}, {Imran}, {Kaaret}, {Karlsson},
  {Kertzman}, {Kieda}, {Krawczynski}, {Krennrich}, {Lang}, {LeBohec},
  {Madhavan}, {Maier}, {Majumdar}, {McArthur}, {McCann}, {Moriarty},
  {Mukherjee}, {Nu{\~n}ez}, {Ong}, {Orr}, {Otte}, {Park}, {Perkins}, {Pichel},
  {Pohl}, {Prokoph}, {Quinn}, {Ragan}, {Reyes}, {Reynolds}, {Roache}, {Rose},
  {Ruppel}, {Saxon}, {Schroedter}, {Sembroski}, {{\c{S}}ent{\"u}rk}, {Skole},
  {Staszak}, {Te{\v{s}}i{\'c}}, {Theiling}, {Thibadeau}, {Tsurusaki}, {Tyler},
  {Varlotta}, {Vassiliev}, {Vincent}, {Vivier}, {Wakely}, {Ward}, {Weekes},
  {Weinstein}, {Weisgarber}, {Williams}, \& {Zitzer}}]{Aliu:2012:ApJ}
{Aliu}, E., {Arlen}, T., {Aune}, T., {et~al.} 2012, \apj, 746, 141,
  \dodoi{10.1088/0004-637X/746/2/141}

\bibitem[{{Asada} \& {Nakamura}(2012)}]{Asada:2012:ApJL}
{Asada}, K., \& {Nakamura}, M. 2012, \apjl, 745, L28,
  \dodoi{10.1088/2041-8205/745/2/L28}

\bibitem[{{Asada} {et~al.}(2016){Asada}, {Nakamura}, \& {Pu}}]{Asada:2016:ApJ}
{Asada}, K., {Nakamura}, M., \& {Pu}, H.-Y. 2016, \apj, 833, 56,
  \dodoi{10.3847/1538-4357/833/1/56}

\bibitem[{{Beskin} {et~al.}(1992){Beskin}, {Istomin}, \& {Parev}}]{bes92}
{Beskin}, V.~S., {Istomin}, Y.~N., \& {Parev}, V.~I. 1992, \sovast, 36, 642

\bibitem[{{Biretta} {et~al.}(1999){Biretta}, {Sparks}, \&
  {Macchetto}}]{Biretta:1999:ApJ}
{Biretta}, J.~A., {Sparks}, W.~B., \& {Macchetto}, F. 1999, \apj, 520, 621,
  \dodoi{10.1086/307499}

\bibitem[{{Blandford} \& {Levinson}(1995)}]{Blandford:1995:ApJ}
{Blandford}, R.~D., \& {Levinson}, A. 1995, \apj, 441, 79,
  \dodoi{10.1086/175338}

\bibitem[{{Blandford} \& {Payne}(1982)}]{blandford:1982MNRAS}
{Blandford}, R.~D., \& {Payne}, D.~G. 1982, \mnras, 199, 883,
  \dodoi{10.1093/mnras/199.4.883}

\bibitem[{{Blandford} \& {Znajek}(1977)}]{bla77}
{Blandford}, R.~D., \& {Znajek}, R.~L. 1977, \mnras, 179, 433,
  \dodoi{10.1093/mnras/179.3.433}

\bibitem[{{Boccardi} {et~al.}(2016){Boccardi}, {Krichbaum}, {Bach}, {Mertens},
  {Ros}, {Alef}, \& {Zensus}}]{Boccardi:2016:A&A}
{Boccardi}, B., {Krichbaum}, T.~P., {Bach}, U., {et~al.} 2016, \aap, 585, A33,
  \dodoi{10.1051/0004-6361/201526985}

\bibitem[{{Bransgrove} {et~al.}(2021){Bransgrove}, {Ripperda}, \&
  {Philippov}}]{Bransgrove:2021:PhRvL}
{Bransgrove}, A., {Ripperda}, B., \& {Philippov}, A. 2021, \prl, 127, 055101,
  \dodoi{10.1103/PhysRevLett.127.055101}

\bibitem[{{Broderick} \& {Loeb}(2009)}]{2009ApJ...697.1164B}
{Broderick}, A.~E., \& {Loeb}, A. 2009, \apj, 697, 1164,
  \dodoi{10.1088/0004-637X/697/2/1164}

\bibitem[{{Broderick} \& {Tchekhovskoy}(2015)}]{Broderick:2015:ApJ}
{Broderick}, A.~E., \& {Tchekhovskoy}, A. 2015, \apj, 809, 97,
  \dodoi{10.1088/0004-637X/809/1/97}

\bibitem[{{Camenzind}(1986)}]{came86b}
{Camenzind}, M. 1986, \aap, 162, 32

\bibitem[{{Celotti} \& {Fabian}(1993)}]{Celotti:1993:MNRAS:264}
{Celotti}, A., \& {Fabian}, A.~C. 1993, \mnras, 264, 228,
  \dodoi{10.1093/mnras/264.1.228}

\bibitem[{{Chen} \& {Yuan}(2020)}]{Chen:2020:ApJ}
{Chen}, A.~Y., \& {Yuan}, Y. 2020, \apj, 895, 121,
  \dodoi{10.3847/1538-4357/ab8c46}

\bibitem[{{Chiueh} {et~al.}(1998){Chiueh}, {Li}, \&
  {Begelman}}]{Chiueh:1998:ApJ}
{Chiueh}, T., {Li}, Z.-Y., \& {Begelman}, M.~C. 1998, \apj, 505, 835,
  \dodoi{10.1086/306209}

\bibitem[{{Crinquand} {et~al.}(2021){Crinquand}, {Cerutti}, {Dubus}, {Parfrey},
  \& {Philippov}}]{Crinquand:2021:AA}
{Crinquand}, B., {Cerutti}, B., {Dubus}, G., {Parfrey}, K., \& {Philippov}, A.
  2021, \aap, 650, A163, \dodoi{10.1051/0004-6361/202040158}

\bibitem[{{Dokuchaev}(2023)}]{Dokuchaev:2023:Astro}
{Dokuchaev}, V.~I. 2023, Astronomy, 2, 141, \dodoi{10.3390/astronomy2030010}

\bibitem[{{Event Horizon Telescope Collaboration} {et~al.}(2019){Event Horizon
  Telescope Collaboration}, {Akiyama}, {Alberdi}, {Alef}, {Asada}, {Azulay},
  {Baczko}, {Ball}, {Balokovi{\'c}}, {Barrett}, {Bintley}, {Blackburn},
  {Boland}, {Bouman}, {Bower}, {Bremer}, {Brinkerink}, {Brissenden}, {Britzen},
  {Broderick}, {Broguiere}, {Bronzwaer}, {Byun}, {Carlstrom}, {Chael}, {Chan},
  {Chatterjee}, {Chatterjee}, {Chen}, {Chen}, {Cho}, {Christian}, {Conway},
  {Cordes}, {Crew}, {Cui}, {Davelaar}, {De Laurentis}, {Deane}, {Dempsey},
  {Desvignes}, {Dexter}, {Doeleman}, {Eatough}, {Falcke}, {Fish}, {Fomalont},
  {Fraga-Encinas}, {Friberg}, {Fromm}, {G{\'o}mez}, {Galison}, {Gammie},
  {Garc{\'\i}a}, {Gentaz}, {Georgiev}, {Goddi}, {Gold}, {Gu}, {Gurwell},
  {Hada}, {Hecht}, {Hesper}, {Ho}, {Ho}, {Honma}, {Huang}, {Huang}, {Hughes},
  {Ikeda}, {Inoue}, {Issaoun}, {James}, {Jannuzi}, {Janssen}, {Jeter}, {Jiang},
  {Johnson}, {Jorstad}, {Jung}, {Karami}, {Karuppusamy}, {Kawashima},
  {Keating}, {Kettenis}, {Kim}, {Kim}, {Kim}, {Kino}, {Koay}, {Koch}, {Koyama},
  {Kramer}, {Kramer}, {Krichbaum}, {Kuo}, {Lauer}, {Lee}, {Li}, {Li},
  {Lindqvist}, {Liu}, {Liuzzo}, {Lo}, {Lobanov}, {Loinard}, {Lonsdale}, {Lu},
  {MacDonald}, {Mao}, {Markoff}, {Marrone}, {Marscher}, {Mart{\'\i}-Vidal},
  {Matsushita}, {Matthews}, {Medeiros}, {Menten}, {Mizuno}, {Mizuno}, {Moran},
  {Moriyama}, {Moscibrodzka}, {M{\"u}ller}, {Nagai}, {Nagar}, {Nakamura},
  {Narayan}, {Narayanan}, {Natarajan}, {Neri}, {Ni}, {Noutsos}, {Okino},
  {Olivares}, {Oyama}, {{\"O}zel}, {Palumbo}, {Patel}, {Pen}, {Pesce},
  {Pi{\'e}tu}, {Plambeck}, {PopStefanija}, {Porth}, {Prather},
  {Preciado-L{\'o}pez}, {Psaltis}, {Pu}, {Ramakrishnan}, {Rao}, {Rawlings},
  {Raymond}, {Rezzolla}, {Ripperda}, {Roelofs}, {Rogers}, {Ros}, {Rose},
  {Roshanineshat}, {Rottmann}, {Roy}, {Ruszczyk}, {Ryan}, {Rygl},
  {S{\'a}nchez}, {S{\'a}nchez-Arguelles}, {Sasada}, {Savolainen}, {Schloerb},
  {Schuster}, {Shao}, {Shen}, {Small}, {Sohn}, {SooHoo}, {Tazaki}, {Tiede},
  {Tilanus}, {Titus}, {Toma}, {Torne}, {Trent}, {Trippe}, {Tsuda}, {van
  Bemmel}, {van Langevelde}, {van Rossum}, {Wagner}, {Wardle}, {Weintroub},
  {Wex}, {Wharton}, {Wielgus}, {Wong}, {Wu}, {Young}, {Young}, {Younsi},
  {Yuan}, {Yuan}, {Zensus}, {Zhao}, {Zhao}, {Zhu}, {Farah}, {Meyer-Zhao},
  {Michalik}, {Nadolski}, {Nishioka}, {Pradel}, {Primiani}, {Souccar},
  {Vertatschitsch}, \&
  {Yamaguchi}}]{EventHorizonTelescopeCollaboration:2019:ApJL}
{Event Horizon Telescope Collaboration}, {Akiyama}, K., {Alberdi}, A., {et~al.}
  2019, \apjl, 875, L6, \dodoi{10.3847/2041-8213/ab1141}

\bibitem[{{Feng} \& {Wu}(2017)}]{Feng:2017:MNRAS}
{Feng}, J., \& {Wu}, Q. 2017, \mnras, 470, 612, \dodoi{10.1093/mnras/stx1283}

\bibitem[{{Gebhardt} {et~al.}(2011){Gebhardt}, {Adams}, {Richstone}, {Lauer},
  {Faber}, {G{\"u}ltekin}, {Murphy}, \& {Tremaine}}]{Gebhardt:2011:ApJ}
{Gebhardt}, K., {Adams}, J., {Richstone}, D., {et~al.} 2011, \apj, 729, 119,
  \dodoi{10.1088/0004-637X/729/2/119}

\bibitem[{{Ginzburg} \& {Syrovatskii}(1965)}]{Ginzburg:1965}
{Ginzburg}, V.~L., \& {Syrovatskii}, S.~I. 1965, \araa, 3, 297,
  \dodoi{10.1146/annurev.aa.03.090165.001501}

\bibitem[{{Giroletti} {et~al.}(2004){Giroletti}, {Giovannini}, {Feretti},
  {Cotton}, {Edwards}, {Lara}, {Marscher}, {Mattox}, {Piner}, \&
  {Venturi}}]{Giroletti:2004:ApJ}
{Giroletti}, M., {Giovannini}, G., {Feretti}, L., {et~al.} 2004, \apj, 600,
  127, \dodoi{10.1086/379663}

\bibitem[{{Hada} {et~al.}(2013){Hada}, {Kino}, {Doi}, {Nagai}, {Honma},
  {Hagiwara}, {Giroletti}, {Giovannini}, \& {Kawaguchi}}]{hada13}
{Hada}, K., {Kino}, M., {Doi}, A., {et~al.} 2013, \apj, 775, 70,
  \dodoi{10.1088/0004-637X/775/1/70}

\bibitem[{{Hada} {et~al.}(2016){Hada}, {Kino}, {Doi}, {Nagai}, {Honma},
  {Akiyama}, {Tazaki}, {Lico}, {Giroletti}, {Giovannini}, {Orienti}, \&
  {Hagiwara}}]{hada16}
---. 2016, \apj, 817, 131, \dodoi{10.3847/0004-637X/817/2/131}

\bibitem[{{Hirotani}(2005)}]{Hirotani:2005:ApJ}
{Hirotani}, K. 2005, \apj, 619, 73, \dodoi{10.1086/426497}

\bibitem[{{Hirotani} \& {Okamoto}(1998)}]{Hirotani:1998:ApJ}
{Hirotani}, K., \& {Okamoto}, I. 1998, \apj, 497, 563, \dodoi{10.1086/305479}

\bibitem[{{Hirotani} \& {Pu}(2016)}]{hiro16a}
{Hirotani}, K., \& {Pu}, H.-Y. 2016, \apj, 818, 50,
  \dodoi{10.3847/0004-637X/818/1/50}

\bibitem[{{Hirotani} {et~al.}(2016){Hirotani}, {Pu}, {Lin}, {Chang}, {Inoue},
  {Kong}, {Matsushita}, \& {Tam}}]{Hirotani:2016:ApJ}
{Hirotani}, K., {Pu}, H.-Y., {Lin}, L. C.-C., {et~al.} 2016, \apj, 833, 142,
  \dodoi{10.3847/1538-4357/833/2/142}

\bibitem[{{Hirotani} {et~al.}(2023){Hirotani}, {Shang}, {Krasnopolsky}, \&
  {Nishikawa}}]{Hirotani:2023:ApJ}
{Hirotani}, K., {Shang}, H., {Krasnopolsky}, R., \& {Nishikawa}, K. 2023, \apj,
  943, 164, \dodoi{10.3847/1538-4357/aca8b0}

\bibitem[{{Hirotani} {et~al.}(1992){Hirotani}, {Takahashi}, {Nitta}, \&
  {Tomimatsu}}]{Hirotani:1992:ApJ}
{Hirotani}, K., {Takahashi}, M., {Nitta}, S.-Y., \& {Tomimatsu}, A. 1992, \apj,
  386, 455, \dodoi{10.1086/171031}

\bibitem[{{Ito} {et~al.}(2008){Ito}, {Kino}, {Kawakatu}, {Isobe}, \&
  {Yamada}}]{Ito:2008:ApJ:685}
{Ito}, H., {Kino}, M., {Kawakatu}, N., {Isobe}, N., \& {Yamada}, S. 2008, \apj,
  685, 828, \dodoi{10.1086/591036}

\bibitem[{{Kellermann} \& {Pauliny-Toth}(1981)}]{Kellermann:1981:ARAA}
{Kellermann}, K.~I., \& {Pauliny-Toth}, I.~I.~K. 1981, \araa, 19, 373,
  \dodoi{10.1146/annurev.aa.19.090181.002105}

\bibitem[{{Kim} {et~al.}(2018){Kim}, {Krichbaum}, {Lu}, {Ros}, {Bach},
  {Bremer}, {de Vicente}, {Lindqvist}, \& {Zensus}}]{Kim:2018:A&A}
{Kim}, J.~Y., {Krichbaum}, T.~P., {Lu}, R.~S., {et~al.} 2018, \aap, 616, A188,
  \dodoi{10.1051/0004-6361/201832921}

\bibitem[{{Kim} {et~al.}(2019){Kim}, {Krichbaum}, {Marscher}, {Jorstad},
  {Agudo}, {Thum}, {Hodgson}, {MacDonald}, {Ros}, {Lu}, {Bremer}, {de Vicente},
  {Lindqvist}, {Trippe}, \& {Zensus}}]{Kim:2019:A&A}
{Kim}, J.~Y., {Krichbaum}, T.~P., {Marscher}, A.~P., {et~al.} 2019, \aap, 622,
  A196, \dodoi{10.1051/0004-6361/201832920}

\bibitem[{{Kisaka} {et~al.}(2020){Kisaka}, {Levinson}, \&
  {Toma}}]{Kisaka:2020:ApJ}
{Kisaka}, S., {Levinson}, A., \& {Toma}, K. 2020, \apj, 902, 80,
  \dodoi{10.3847/1538-4357/abb46c}

\bibitem[{{Koide} {et~al.}(2002){Koide}, {Shibata}, {Kudoh}, \&
  {Meier}}]{Koide:2002:Sci}
{Koide}, S., {Shibata}, K., {Kudoh}, T., \& {Meier}, D.~L. 2002, Science, 295,
  1688, \dodoi{10.1126/science.1068240}

\bibitem[{{Komissarov}(2005)}]{komissarov:2005MNRAS}
{Komissarov}, S.~S. 2005, \mnras, 359, 801,
  \dodoi{10.1111/j.1365-2966.2005.08974.x}

\bibitem[{{Kovalev} {et~al.}(2007){Kovalev}, {Lister}, {Homan}, \&
  {Kellermann}}]{Kovalev:2007:ApJL}
{Kovalev}, Y.~Y., {Lister}, M.~L., {Homan}, D.~C., \& {Kellermann}, K.~I. 2007,
  \apjl, 668, L27, \dodoi{10.1086/522603}

\bibitem[{{Le Roux}(1961)}]{LeRoux:1961}
{Le Roux}, E. 1961, Annales d'Astrophysique, 24, 71

\bibitem[{{Leung} {et~al.}(2011){Leung}, {Gammie}, \& {Noble}}]{Leung:2011:ApJ}
{Leung}, P.~K., {Gammie}, C.~F., \& {Noble}, S.~C. 2011, \apj, 737, 21,
  \dodoi{10.1088/0004-637X/737/1/21}

\bibitem[{{Levinson} \& {Blandford}(1995)}]{Levinson:1995:ApJ}
{Levinson}, A., \& {Blandford}, R. 1995, \apj, 449, 86, \dodoi{10.1086/176034}

\bibitem[{{Levinson} \& {Blandford}(1996)}]{Levinson:1996:ApJL}
---. 1996, \apjl, 456, L29, \dodoi{10.1086/309851}

\bibitem[{{Levinson} \& {Rieger}(2011)}]{levi11}
{Levinson}, A., \& {Rieger}, F. 2011, \apj, 730, 123,
  \dodoi{10.1088/0004-637X/730/2/123}

\bibitem[{{Li} {et~al.}(2009){Li}, {Yuan}, {Wang}, {Wang}, \&
  {Zhang}}]{Li:2009:ApJ}
{Li}, Y.-R., {Yuan}, Y.-F., {Wang}, J.-M., {Wang}, J.-C., \& {Zhang}, S. 2009,
  \apj, 699, 513, \dodoi{10.1088/0004-637X/699/1/513}

\bibitem[{{Li} {et~al.}(1992){Li}, {Chiueh}, \& {Begelman}}]{Li:1992:ApJ}
{Li}, Z.-Y., {Chiueh}, T., \& {Begelman}, M.~C. 1992, \apj, 394, 459,
  \dodoi{10.1086/171597}

\bibitem[{{Liepold} {et~al.}(2023){Liepold}, {Ma}, \&
  {Walsh}}]{Liepold:2023:ApJL}
{Liepold}, E.~R., {Ma}, C.-P., \& {Walsh}, J.~L. 2023, \apjl, 945, L35,
  \dodoi{10.3847/2041-8213/acbbcf}

\bibitem[{{Lu} {et~al.}(2023){Lu}, {Asada}, {Krichbaum}, {Park}, {Tazaki},
  {Pu}, {Nakamura}, {Lobanov}, {Hada}, {Akiyama}, {Kim}, {Marti-Vidal},
  {G{\'o}mez}, {Kawashima}, {Yuan}, {Ros}, {Alef}, {Britzen}, {Bremer},
  {Broderick}, {Doi}, {Giovannini}, {Giroletti}, {Ho}, {Honma}, {Hughes},
  {Inoue}, {Jiang}, {Kino}, {Koyama}, {Lindqvist}, {Liu}, {Marscher},
  {Matsushita}, {Nagai}, {Rottmann}, {Savolainen}, {Schuster}, {Shen}, {de
  Vicente}, {Walker}, {Yang}, {Zensus}, {Algaba}, {Allardi}, {Bach},
  {Berthold}, {Bintley}, {Byun}, {Casadio}, {Chang}, {Chang}, {Chang}, {Chen},
  {Chen}, {Chilson}, {Chuter}, {Conway}, {Crew}, {Dempsey}, {Dornbusch},
  {Faber}, {Friberg}, {Garc{\'\i}a}, {Garrido}, {Han}, {Han}, {Hasegawa},
  {Herrero-Illana}, {Huang}, {Huang}, {Impellizzeri}, {Jiang}, {Jinchi},
  {Jung}, {Kallunki}, {Kirves}, {Kimura}, {Koay}, {Koch}, {Kramer}, {Kraus},
  {Kubo}, {Kuo}, {Li}, {Lin}, {Liu}, {Liu}, {Lo}, {Lu}, {MacDonald},
  {Martin-Cocher}, {Messias}, {Meyer-Zhao}, {Minter}, {Nair}, {Nishioka},
  {Norton}, {Nystrom}, {Ogawa}, {Oshiro}, {Patel}, {Pen}, {Pidopryhora},
  {Pradel}, {Raffin}, {Rao}, {Ruiz}, {Sanchez}, {Shaw}, {Snow}, {Sridharan},
  {Srinivasan}, {Tercero}, {Torne}, {Traianou}, {Wagner}, {Walther}, {Wei},
  {Yang}, \& {Yu}}]{Lu:2023:Natur}
{Lu}, R.-S., {Asada}, K., {Krichbaum}, T.~P., {et~al.} 2023, \nat, 616, 686,
  \dodoi{10.1038/s41586-023-05843-w}

\bibitem[{{Matsumoto} \& {Masada}(2013)}]{Matsumoto:2013:ApJL}
{Matsumoto}, J., \& {Masada}, Y. 2013, \apjl, 772, L1,
  \dodoi{10.1088/2041-8205/772/1/L1}

\bibitem[{{McKinney} \& {Gammie}(2004)}]{mckinney:2004ApJ}
{McKinney}, J.~C., \& {Gammie}, C.~F. 2004, \apj, 611, 977,
  \dodoi{10.1086/422244}

\bibitem[{{McKinney} {et~al.}(2012){McKinney}, {Tchekhovskoy}, \& {Bland
  ford}}]{McKinney:2012:MNRAS}
{McKinney}, J.~C., {Tchekhovskoy}, A., \& {Bland ford}, R.~D. 2012, \mnras,
  423, 3083, \dodoi{10.1111/j.1365-2966.2012.21074.x}

\bibitem[{{Mertens} {et~al.}(2016){Mertens}, {Lobanov}, {Walker}, \&
  {Hardee}}]{Mertens:2016:AA}
{Mertens}, F., {Lobanov}, A.~P., {Walker}, R.~C., \& {Hardee}, P.~E. 2016,
  \aap, 595, A54, \dodoi{10.1051/0004-6361/201628829}

\bibitem[{{Miley}(1980)}]{Miley:1980:ARAA}
{Miley}, G. 1980, \araa, 18, 165, \dodoi{10.1146/annurev.aa.18.090180.001121}

\bibitem[{{Nagai} {et~al.}(2014){Nagai}, {Haga}, {Giovannini}, {Doi},
  {Orienti}, {D'Ammando}, {Kino}, {Nakamura}, {Asada}, {Hada}, \&
  {Giroletti}}]{Nagai:2014:ApJ}
{Nagai}, H., {Haga}, T., {Giovannini}, G., {et~al.} 2014, \apj, 785, 53,
  \dodoi{10.1088/0004-637X/785/1/53}

\bibitem[{{Nakamura} {et~al.}(2018){Nakamura}, {Asada}, {Hada}, {Pu}, {Noble},
  {Tseng}, {Toma}, {Kino}, {Nagai}, {Takahashi}, {Algaba}, {Orienti},
  {Akiyama}, {Doi}, {Giovannini}, {Giroletti}, {Honma}, {Koyama}, {Lico},
  {Niinuma}, \& {Tazaki}}]{Nakamura:2018:ApJ}
{Nakamura}, M., {Asada}, K., {Hada}, K., {et~al.} 2018, \apj, 868, 146,
  \dodoi{10.3847/1538-4357/aaeb2d}

\bibitem[{{Neronov} \& {Aharonian}(2007)}]{nero07}
{Neronov}, A., \& {Aharonian}, F.~A. 2007, \apj, 671, 85,
  \dodoi{10.1086/522199}

\bibitem[{{Parfrey} {et~al.}(2015){Parfrey}, {Giannios}, \&
  {Beloborodov}}]{Parfrey:2015:MNRAS}
{Parfrey}, K., {Giannios}, D., \& {Beloborodov}, A.~M. 2015, \mnras, 446, L61,
  \dodoi{10.1093/mnrasl/slu162}

\bibitem[{{Parfrey} {et~al.}(2019){Parfrey}, {Philippov}, \&
  {Cerutti}}]{Parfrey:2019:PhRvL}
{Parfrey}, K., {Philippov}, A., \& {Cerutti}, B. 2019, \prl, 122, 035101,
  \dodoi{10.1103/PhysRevLett.122.035101}

\bibitem[{{Perlman} {et~al.}(2011){Perlman}, {Adams}, {Cara}, {Bourque},
  {Harris}, {Madrid}, {Simons}, {Clausen-Brown}, {Cheung}, {Stawarz},
  {Georganopoulos}, {Sparks}, \& {Biretta}}]{Perlman:2011:ApJ}
{Perlman}, E.~S., {Adams}, S.~C., {Cara}, M., {et~al.} 2011, \apj, 743, 119,
  \dodoi{10.1088/0004-637X/743/2/119}

\bibitem[{{Qian} {et~al.}(2018){Qian}, {Fendt}, \& {Vourellis}}]{Qian:2018ApJ}
{Qian}, Q., {Fendt}, C., \& {Vourellis}, C. 2018, \apj, 859, 28,
  \dodoi{10.3847/1538-4357/aabd36}

\bibitem[{{Rieger} \& {Aharonian}(2008)}]{Rieger:2008:A&A}
{Rieger}, F.~M., \& {Aharonian}, F.~A. 2008, \aap, 479, L5,
  \dodoi{10.1051/0004-6361:20078706}

\bibitem[{{Rybicki} \& {Lightman}(1979)}]{Rybicki:1979:rpa}
{Rybicki}, G.~B., \& {Lightman}, A.~P. 1979, {Radiative processes in
  astrophysics}

\bibitem[{{Song} {et~al.}(2017){Song}, {Pu}, {Hirotani}, {Matsushita}, {Kong},
  \& {Chang}}]{Song:2017:MNRAS}
{Song}, Y., {Pu}, H.-Y., {Hirotani}, K., {et~al.} 2017, \mnras, 471, L135,
  \dodoi{10.1093/mnrasl/slx119}

\bibitem[{{Takahashi} {et~al.}(2018){Takahashi}, {Toma}, {Kino}, {Nakamura}, \&
  {Hada}}]{Takahashi:2018:ApJ}
{Takahashi}, K., {Toma}, K., {Kino}, M., {Nakamura}, M., \& {Hada}, K. 2018,
  \apj, 868, 82, \dodoi{10.3847/1538-4357/aae832}

\bibitem[{{Takahashi} {et~al.}(1990){Takahashi}, {Nitta}, {Tatematsu}, \&
  {Tomimatsu}}]{takahashi:1990ApJ}
{Takahashi}, M., {Nitta}, S., {Tatematsu}, Y., \& {Tomimatsu}, A. 1990, \apj,
  363, 206, \dodoi{10.1086/169331}

\bibitem[{{Tchekhovskoy} {et~al.}(2010){Tchekhovskoy}, {Narayan}, \&
  {McKinney}}]{Tchekhovskoy:2010:ApJ}
{Tchekhovskoy}, A., {Narayan}, R., \& {McKinney}, J.~C. 2010, \apj, 711, 50,
  \dodoi{10.1088/0004-637X/711/1/50}

\bibitem[{{Tchekhovskoy} {et~al.}(2011){Tchekhovskoy}, {Narayan}, \&
  {McKinney}}]{Tchekhovskoy:2011:MNRAS}
---. 2011, \mnras, 418, L79, \dodoi{10.1111/j.1745-3933.2011.01147.x}

\bibitem[{{Toma} \& {Takahara}(2013)}]{toma2013PTEP}
{Toma}, K., \& {Takahara}, F. 2013, Progress of Theoretical and Experimental
  Physics, 2013, 083E02, \dodoi{10.1093/ptep/ptt058}

\bibitem[{{Tomimatsu} \& {Takahashi}(2003)}]{Tomimatsu:2003:ApJ}
{Tomimatsu}, A., \& {Takahashi}, M. 2003, \apj, 592, 321,
  \dodoi{10.1086/375579}

\bibitem[{{Vlahakis} \& {K{\"o}nigl}(2003)}]{Vlahakis:2003:ApJ}
{Vlahakis}, N., \& {K{\"o}nigl}, A. 2003, \apj, 596, 1080,
  \dodoi{10.1086/378226}

\bibitem[{{Wardzi{\'n}ski} \& {Zdziarski}(2000)}]{Wardzinski:2000:MNRAS}
{Wardzi{\'n}ski}, G., \& {Zdziarski}, A.~A. 2000, \mnras, 314, 183,
  \dodoi{10.1046/j.1365-8711.2000.03297.x}

\bibitem[{{Znajek}(1977)}]{Znajek:1977:MNRAS}
{Znajek}, R.~L. 1977, \mnras, 179, 457, \dodoi{10.1093/mnras/179.3.457}

\end{thebibliography}

\end{document}